\documentclass[12pt,thmsa]{article}
\usepackage{amssymb}

\usepackage{epsfig}
\usepackage{rotating}
\usepackage{portland}
\usepackage{lscape}
\usepackage{sw20lart}
\usepackage[dvips]{tcigraph}

\input{tcilatex}
\begin{document}

\title{Solar and Gravitation Nonlinearity contributions to Mercury perihelion
Advance}
\author{Sophie Pireaux\thanks{%
D\'{e}partement de Physique Th\'{e}orique et Math\'{e}matique (FYMA), B301%
\newline
Universit\'{e} de Louvain La Neuve,\newline
2 Chemin du Cyclotron,\newline
1348 Louvain La Neuve, BELGIUM\newline
TEL: ++32(0)10/47 32 85 and FAX: ++32(0)10/47 24 14\newline
E-mail: pireaux@fyma.ucl.ac.be}, Jean-Pierre Rozelot and St\'{e}phany Godier 
\thanks{%
CERGA Department,\newline
Observtoire de la C\^{o}te d'Azur,\newline
Avenue Copernic,\newline
Grasse, FRANCE\newline
Tel: ++33(0)4/93 40 53 54 and Fax : ++33(0)4/93 40 53 33\newline
E-mail: St\'{e}phany.Godier@obs-azur.fr }}

\rotdriver{dvips}

\begin{center}
{\Large Solar quadrupole moment and purely relativistic gravitation
contributions to Mercury's perihelion Advance}\bigskip

Sophie Pireaux\footnote{{\tiny D\'{e}partement de Physique Th\'{e}orique et
Math\'{e}matique (FYMA), B301\newline
Universit\'{e} de Louvain La Neuve, 2 Chemin du Cyclotron, 1348 Louvain La
Neuve, BELGIUM\newline
TEL: ++32(0)10/47 32 85 and FAX: ++32(0)10/47 24 14\quad E-mail:
pireaux@fyma.ucl.ac.be\medskip }}, Jean-Pierre Rozelot and St\'{e}phany
Godier\footnote{{\tiny CERGA Department,\newline
Observatoire de la C\^{o}te d'Azur, Avenue Copernic, Grasse, FRANCE\newline
Tel: ++33(0)4/93 40 53 54 and Fax : ++33(0)4/93 40 53 33\quad E-mail:
rozelot@obs-azur.fr}\newline
{\tiny \medskip and St\'{e}phany.Godier@obs-azur.fr }}

Received 3rd September 2001.\bigskip 
\end{center}

{\footnotesize ABSTRACT: The perihelion advance of the orbit of Mercury has
long been one of the observational cornerstones for testing General
Relativity (G.R.).}

{\footnotesize The main goal of this paper is to discuss how, presently,
observational and theoretical constraints may challenge Einstein's theory of
gravitation characterized by }$\beta =\gamma =1${\footnotesize . To achieve
this purpose, we will first recall the experimental constraints upon the
Eddington-Robertson parameters }$\gamma ${\footnotesize , }$\beta $%
{\footnotesize \ and the observational bounds for the perihelion advance of
Mercury, }$\Delta \omega _{obs}${\footnotesize .\newline
A second point will address the values given, up to now, to the solar
quadrupole moment by several authors. Then, we will briefly comment why we
use a recent theoretical determination of the solar quadrupole moment, }$%
J_{2}=(2.0\pm 0.4)\ 10^{-7}${\footnotesize , which takes into account both
surfacic and internal differential rotation, in order to compute the solar
contribution to Mercury's perihelion advance.\newline
Further on, combining bounds on }$\gamma ${\footnotesize \ and }$J_{2}$%
{\footnotesize \ contributions, and taking into account the observational
data range for }$\Delta \omega _{obs}${\footnotesize , we will be able to
give a range of values for }$\beta ${\footnotesize .}

{\footnotesize Alternatively, taking into account the observed value of }$%
\Delta \omega _{obs}${\footnotesize , one can deduce a dynamical estimation
of }$J_{2}${\footnotesize \ in the setting of G.R. This point is important
as it provides a solar model independent estimation that can be confronted
with other determinations of }$J_{2}${\footnotesize \ based upon solar
theory and solar observations (oscillation data, oblateness...).}

{\footnotesize Finally, a glimpse at future satellite experiments will help
us to understand how stronger constraints upon the parameter space (}$\gamma 
${\footnotesize , }$\beta ${\footnotesize ,\ }$J_{2}${\footnotesize ) as
well as a separation of the two contributions (from the quadrupole moment, }$%
J_{2}${\footnotesize , or purely relativistic, }$2\alpha ^{2}+2\alpha \gamma
-\beta ${\footnotesize ) might be expected in the future.}


\section{INTRODUCTION}

\qquad The solar quadrupole moment, $J_{2}$, is one of the fundamental
figures in solar physics. It provides informations on the distortion of the
effective solar potential, $J_{2}$ being the first perturbation coefficient
to a pure spherically symmetric gravitational field: 
\[
\Phi (r,\theta )=-\frac{GM}{r}\left[ 1-\sum_{n=1}^{\infty }\left( \frac{R_{s}%
}{r}\right) ^{2}J_{n}\ P_{n}(\cos \theta )\right] 
\]
where $\Phi $ is the solar component of the gravitational potential outside
the Sun, in polar coordinates ($r,\theta ,\phi $) with respect to the Sun's
rotation axis; $P_{n}$ are Legendre functions of degree $n$. The
coefficients $J_{n}$ are thus directly related to the distorted shape of the
Sun; for instance, for $n=2$, $J_{2}\neq 0\ $ is an indicator of the
oblateness\footnote{%
Notice that $J_{2}=-c_{02}$, where $c_{02}$ is the second spherical harmonic
coefficient.}.\newline
Concerning the Sun, $J_{2}$, which is the most important term, should be
used as a constraint in the computation of solar models, as the asphericity
is a probe to test the solar interior.\newline
Further, detection of long term changes in the solar figure (as there is
some evidence for $J_{2}$ to vary with time) are intended; those have been
postulated to act as a potential gravitational reservoir that can be a
source of solar luminosity variations, which in turn, could have significant
effects on the climate of the Earth (\cite{Sofia 1979 J2 and climate}, \cite
{Rozelot 2001a J2 and climate}).

Today, the quadrupolar moment is also a non negligeable quantity in
computing the relativistic motion of planets.\newline
The first time that the solar quadrupole moment was associated with
gravitational motion of Mercury is in 1885, when Newcomb attempted to
account for the anomalous perihelion advance of Mercury with a modified
gravitational field, manifested by an oblateness $\Delta r$ \cite{Newcomb
1895-1898 Perihelion advance and J2}. Indeed, in 1859, Le Verrier had
observed a deviation of Mercury's orbit from Newtonian's predictions, that
could not be due to the presence of known planets. But, the difference
between the equatorial and polar diameters of the Sun of $500$ arc ms, as
advocated by Newcomb, was soon ruled out by solar observations. And
Einstein's new theory of gravitation, General Relativity, could account for 
\textit{almost all} the observed perihelion advance.\newline
So, Mercury's perihelion advance readily became one of the cornerstones for
testing General Relativity; even though, now, a contribution to the
perihelion shift from the solar figure (though less important than first
suggested by Newcomb) can not be discarded.

Mercury is the inner most of the four terrestrial planets in the Solar
System, moving with a high velocity in the Sun's gravitational field. Only
comets and asteroids approach the Sun closer at perihelion. This why the
``Mercury lab'' and minor planets too (see section \ref{perihelion
precession of minor planets}) offer unique possibilities for testing G.R.
and exploring the limits of alternative theories of gravitation with an
interesting accuracy.\newline
However, the perihelion shift of planets, and hence Mercury, can not be
measured directly because the perihelion is a Keplerian element whereas the
motions of the planets are not exactly Keplerian due to mutual gravitational
interactions and figure effects. So, only an indirect determination can be
done. One can proceed as follows. The motions of planets, from numerically
integrated ephemeris, are computed over an interval of time. The time
evolution of osculating elements is then plot and a polynomial fit of the
parameters gives the rate of the perihelion advance. If one repeats this
procedure in the classical Newtonian limit, one gets another set of rates.
The difference between the two computations, and taking into account the
constant general precession of the equinoxes, gives the combined effect due
to relativistic gravitation and the Sun's quadrupole moment, $\Delta \omega
_{obs}$.\newline
Nevertheless, $\Delta \omega _{obs}$ depends on how the perturbation
elements (for example the slow motion of the ecliptic) are taken into
account in the computations \cite{Narlikar 1985 N-body calculation}; it also
depends on the precision and on the data set selected from the radar data 
\cite{Rana 1987 motion of the node} which provide the core of the
ephemerides computation.\newline
Furthermore, in the article \cite{Pitjeva1993 Mercury toppography}, the
author shows that the topographic features of Mercury's surface influence
the results on the perihelion advance inferred from Mercury radar
observations.\newline
These are the main reasons for which the range of Mercury's perihelion
advance deduced from the radar data remains of a great amplitude (see table
1, section \ref{Tables}).\textbf{\newline
}M. Standish, \cite{Standish 2000 private communication}, has applied a
method analoguous the above mentioned method; integrating equations over
four centuries, 1800-2200, with and without the relativistic contribution
(the second integration was done by simply replacing the speed of light with
a very large value). He then computed the perihelion of Mercury from both of
the runs, one point every 400 days, and differentiated the values of the
perihelion at each time-point. After fitting a linear function to the
differences, the resulting slope from the figure is: $42.980\pm 0.002$
arcsec/cy. Both integrations assumed $J_{2}=(2.0\pm 0.4)\ 10^{-7}$, hence
the estimation of the perihelion advance given by M. Standish represents
solely the purely relativistic contribution.\newline
At such a level of precision, this solution would scarcely change with
future ephemeris improvements (or with another set of ephemerides
calculations, such as those given by the Bureau des Longitudes, since 1889 
\cite{Bureau des Longitudes 1989}).

In the following second section, we will describe how, using the most
accurate theoretical value for $J_{2}$ , the observed perihelion advance can
lead to constraints upon the parameters ($\beta $, $\gamma $) which describe
a generic, metric and conservative theory of gravitation.\newline
In section 3, we will see how, in the setting of General Relativity, $\Delta
\omega _{obs}$ can be used to provide a dynamical, solar model independent,
estimation of $J_{2}$. This dynamical value can be confronted with that
derived from direct measurements of the solar oblateness or indirect ones
coming from helioseismology, which are solar model dependent.\newline
Finally, we will give, in section 4, an overview of future satellite
experiments that might be expected to put stronger constraints upon the
parameter space ( $\beta $, $\gamma $, $J_{2}$) in the future.\newline
Throughout this article, we will refer to estimated values of $\Delta \omega
_{obs}$ and $J_{2}$ from different authors and sources. Those are listed in
tables at the end of this review, along with the figures, in section \ref
{Tables}.

\section{CONSTRAINTS UPON GRAVITATION THEORIES}

\subsection{The relativistic advance of the perihelion of Mercury}

\subsubsection{The purely relativistic effect.}

\qquad Once correcting for the perturbation due to the general precession of
the equinoxes ($\sim 5000$ (arcsec/cy)) and for the perturbations due to
other planets (computed numerically with a Newtonian N-body model: $\sim 280$%
(arcsec/cy) from Venus, $\sim 150$ (arcsec/cy) from Jupiter and $\sim 100$
(arcsec/cy) from the rest), the advance (in regards to the classical
Keplerian prediction) of the perihelion of Mercury is a combination of a
purely relativistic effect and a contribution from the Sun's quadrupole
moment.\newline
It is given by the following general expression\footnote{%
In some references, the coefficient of the term containing the contribution
of the orbit's inclination is improperly written.}: 
\begin{equation}
\begin{array}{l}
\Delta \omega =\Delta \omega _{0\ GR}\ \delta \qquad \text{(rad/revolution)}
\\ 
\text{with } 
\begin{array}[t]{l}
\Delta \omega _{0\ GR}\equiv \frac{3\pi R}{\alpha \ a\ \left( 1-e^{2}\right) 
} \\ 
\delta \equiv \left[ \frac{1}{3}\left( 2\alpha ^{2}+2\alpha \gamma -\beta
\right) -\frac{R_{s}^{2}}{R\ \alpha \ a\left( 1-e^{2}\right) }\ J_{2}\
\left( 3\sin ^{2}i-1\right) \right]
\end{array}
\end{array}
\label{perihelion_advance}
\end{equation}

and where the following parameters\footnote{%
Notice that the value of the Schwarzschild radius of the Sun is more
accurate than the separate values of the gravitational constant, $G$, and
the Sun's mass, $M_{s}$.\medskip} are\medskip 
\begin{eqnarray}
&&R\ \text{,\quad } 
\begin{array}[t]{l}
\text{the Schwarzschild radius of the Sun, }\frac{2GM_{s}}{c^{2}}\text{,} \\ 
\text{given in \cite{European Physical Journal 2000};}
\end{array}
\label{perihelion_advance_parameter_R} \\
&&M_{s}\ \text{,\quad the Sun's mass, given in \cite{Allen 2000};}
\label{perihelion_advance_parameter_Ms} \\
&&R_{s}\ \text{,\quad the Sun's radius, given in \cite{European Physical
Journal 2000};}  \label{perihelion_advance_parameter_Rs} \\
&&J_{2}\ \text{,\quad } 
\begin{array}[t]{l}
\text{the quadrupole moment of the Sun for which we take} \\ 
\text{the theoretical value of }\left( 2.0\pm 0.4\right) \ 10^{-7}\text{, in 
\cite{Godier-Rozelot 2000 Solar Oblateness};}
\end{array}
\label{perihelion_advance_parameter_J2} \\
&&a\ \text{,\quad the semi-major axis of Mercury's orbit, in \cite{Allen
2000};}  \label{perihelion_advance_parameter_a} \\
&&e\ \text{,\quad the exentricity of Mercury's orbit, in \cite{Allen 2000};}
\label{perihelion_advance_parameter_e} \\
&&i\text{ ,\quad the inclination of Mercury's orbit, in \cite{Allen 2000};}
\label{perihelion_advance_parameter_i}
\end{eqnarray}
\medskip

Notice that formula (\ref{perihelion_advance}) is only valid for fully
conservative theories. If it is not the case, the complete expression is
recovered with the following change 
\[
\delta =\left[ 
\begin{array}{l}
\frac{1}{3}\left( 2\alpha ^{2}+2\alpha \gamma -\beta \right) \\ 
-\frac{R_{s}^{2}}{R\ \alpha \ a\left( 1-e^{2}\right) }\ J_{2}\ \left( 3\sin
^{2}i-1\right) \\ 
+\frac{1}{6}\left( 2\alpha _{1}-\alpha _{2}+\alpha _{3}+2\zeta _{2}\right) 
\frac{M_{s}M_{M}}{\left( M_{s}+M_{M}\right) ^{2}}
\end{array}
\right] 
\]
where ``$M_{M}$'' is the mass of Mercury; ``$\alpha _{1}$'', ``$\alpha _{2}$%
'', ``$\alpha _{3}$'' parametrize preferred-frame effects; and ``$\zeta _{2}$%
'', ``$\alpha _{3}$'', a violation of the conservation of the total momentum.%
\newline
But the extra term is nevertheless negligeable because it is proportional to 
$\frac{M_{s}M_{M}}{\left( M_{s}+M_{M}\right) ^{2}}\sim \frac{M_{M}}{M_{s}}%
\sim 2\ 10^{-7}$ (see \cite{Will 1993 Theory and experiment}), and thus
negligeable in regards to the first one (of the order of unity) or to the
second one (of the order of $10^{-4}$).\bigskip

``$\alpha $'', ``$\beta $'' and ``$\gamma $'', refer to the
Eddington-Robertson parameters of the Parametrized Post-Newtonian (P.P.N.)
formalism, describing a fully conservative relativistic theory of
gravitation.\newline
``$\alpha $'' describes the weak equivalence principle; ``$\beta $'' is the
amount of non-linearity in the superposition law of gravity; and ``$\gamma $%
'' characterizes the amount of space curvature produced by unit rest mass.%
\newline
The P.P.N. parameters also cover the particular case of Einstein's theory of
gravitation, General Relativity, characterized by $\alpha =\beta =\gamma =1$.

\subsubsection{Constraints upon the Eddington-Robertson parameters.}

\qquad First of all, the parameter $\alpha $ is set to unity for any theory
that respects the weak equivalence principle, well tested (the difference
between the acceleration towards the Earth of two test-masses of different
composition, relative to the sum of those accelerations, is inferior to $%
\sim 10^{-14}$. See \cite{Will 2001 Theory and experiment update}).\newline
Note that the Microscope Mission, selected by the French agency C.N.E.S. and
scheduled for launch by 2004, has for scientific objective to test the
equivalence principle up to an accuracy of $10^{-15}$, using its well known
manifestation, the universality of free fall \cite{Tobout 2000 Microscope
Mission}.

Secondly, it is light deflection experiments (measuring the combination $%
\frac{\alpha +\gamma }{2}$) that provide so far the best constrains on $%
\gamma $ \cite{Lebach 1995 Light deflection}\footnote{%
According to other authors, \cite{Robertson 1991 VLBI measurments of gamma},
the value of this parameter deduced from V.L.B.I. measurements is $\gamma
=1.0002$\ $\pm 0.00096$; whilts Eubanks et al., as quoted by \cite{Will 2001
Theory and experiment update}, give $\frac{1+\gamma }{2}=0.99992$\ $\pm
0.00014$ (not yet published).}: 
\begin{equation}
\gamma =0.9996\pm 0.0017  \label{estimated value of gamma from L.L.R. data}
\end{equation}

But there is, presently, no independent determination of the parameter $%
\beta $, which appears either in the combination $2\alpha ^{2}+2\alpha
\gamma -\beta $ characterizing the perihelion advance, or in the Nordtvedt
effect $4\beta -\gamma -3\equiv \eta $ (see section \ref{Alternative
theories to G.R, beta different from unity}).

\subsection{Theoretical solar quadrupole moment contribution}

\subsubsection{The question of the accurate determination of $J_{2}$.}

\label{controversies on J2 determination}\qquad The evaluation of the solar
quadrupole moment, $J_{2}$, still faces some controversy: on one side, the
theoretical values strongly depend on the solar model used, whereas accurate
measurements are very difficult to obtain from observations.

Concerning this last point, let us for example recall some problems:\newline
(1) the real differences of brightness of the solar limb dependency on the
latitude; influence of faculae, sunspots and magnetic fields; correlatively,
real effects due to latitudinal variation of the solar limb darkening
function;\newline
(2) the questioned solar activity (solar cycle) dependency of the Sun's
oblateness. These variations were first conjectured by Dicke et al. in 1985 
\newline
\cite{Dicke 1987 Solar oblateness variable?}. Observations at the Pic du
Midi Observatory (France) from 1993 till 2000 seem to confirm a faint
variability reported in previous observations made in 1983-1984.
Nevertheless, the amplitude of the observed variations does not exceed $%
0.02"-0.04"$ over 20 years. (From \cite{Kuhn 1988 Variations of J2}. See
also in \cite{Rozelot 1996 Measure of sun's changing sizes}, fig. 1 and 2;
in \cite{Rozelot 1996 Measure of solar cycle dependency}, and in \cite
{Rozelot 1997 Upper bound for J2} where the authors derive from all the
available data a maximum value of $J_{2}$ of $1\ 10^{-5}$ and an average
value of $\left( 3.64\pm 2.84\right) \ 10^{-6}$);\newline
(3) and the difficulty to calibrate ground data in regards to atmospheric
disturbances (local atmospheric refractive indexes and distortions due to
atmospheric waves).

Space experiments have been suggested in order to solve those problems;
however, first results obtained from the SoHO mission, \cite{Kuhn 1998
Comparison between SoHO and ground-based results on the oblateness}, have
established a good concordance with ground-based observations of the
oblateness (and thus $J_{2}$). Further comments are found in section \ref
{Picard Satellite}.

To illustrate those difficulties, we give a compilation of the main
determinations of $J_{2}$, based on observations and solar theory, in
addition to the main critics to the method used (see table 2 and figure 2,
section \ref{Tables}). A more detailed historical review can be found in 
\cite{Rozelot 1996 Measure of sun's changing sizes} or in \cite{Rozelot 1997
Upper bound for J2}. Remark that early estimations of $J_{2}$, before 1967,
using an heliometer of photographic plates, often erroneously predicted a
prolate Sun (see the second table in \cite{Wittmann 1987 Solar diameter and
its variability}).

In this context, we see that a dynamical determination of $J_{2}$, using the
perihelion shift of Mercury, is interesting as it might be confronted to
those derived from solar model dependent values of the oblateness (see
section \ref{inferring_J2}).

\subsubsection{The adopted theoretical value of $J_{2}$.}

\qquad \label{adopted theoretical value of J2}The theoretical value of $%
J_{2} $, used in this article, has been deduced from a recent work, where
the authors have applied a ``differential theory'' to a solar stratified
model, taking into account the latitudinal differential rotation. The result
is a determination of $J_{2}$ as $\left( 1.60\pm 0.04\right) \ 10^{-7}$ at
the surface of the Sun (see \cite{Godier-Rozelot 1999a rotational potential}
and \cite{Godier-Rozelot 2000 Solar Oblateness}).\newline
The value obtained is in agreement with those calculated by Patern\`{o}%
\newline
\cite{Paterno1996 Sun rotation}, $J_{2}=\left( 2.22\pm 0.1\right) \ 10^{-7}$%
, and Pijpers \cite{Pijpers 1998 quadrupole moment},\newline
$J_{2}=\left( 2.18\pm 0.06\right) \ 10^{-7}$, using in their computations
the inversion techniques applied to helioseismology.\newline
The slight difference between these values and those of Godier/Rozelot comes
mainly from the incertitudes on the solar rotation data due to the
analytical rotation law adopted by \cite{Godier-Rozelot 1999b Relationship
J2 and layers of the Sun} (which gives a low velocity rate at the equator a
bit lower than what is currently observed). But this difference does not
question the order of magnitude\footnote{%
Excluding the unacceptable estimations, that till recently, reported $J_{2}$
to be as large as $10^{-5}$ (see table 2), an order of magnitude larger than
the theoretical upper limit allowed by lunar librations \cite{Rozelot 1998
new results on solar oblateness}.\medskip}, $10^{-4}$, of the solar
contribution in Mercury's perihelion advance. This is why we have admitted
the theoretical range $(2.0\pm 0.4)\ 10^{-7}$ for $J_{2}$.\newline
This value can be confronted to the one given by other authors in table 2 or
figure 2 (section \ref{Tables}).

\subsubsection{G.R.'s prediction with and without the quadrupole moment
contribution.}

\qquad Using the values of the parameters given in the appropriate
references (see (\ref{perihelion_advance_parameter_R}), (\ref
{perihelion_advance_parameter_Ms}), (\ref{perihelion_advance_parameter_Rs}),
(\ref{perihelion_advance_parameter_J2}), (\ref
{perihelion_advance_parameter_a}) and (\ref{perihelion_advance_parameter_e}%
)) in (\ref{perihelion_advance}), plus the value of the period of Mercury's
orbit given in \cite{Allen 2000}, one finds 
\begin{equation}
\Delta \omega _{0\ GR}=\frac{6\pi GM_{s}}{a\left( 1-e^{2}\right) c^{2}}=42.98%
\emph{1}\qquad \text{(arcsec/cy)}  \label{perihelion_advance_GR0}
\end{equation}

for which the accuracy is on the last digit.\newline
This is the prediction of the perihelion shift of Mercury in the setting of
G.R. theory, but omitting the contribution of $J_{2}$.\newline
This raw value is excluded by the last observational data given by \cite
{Anderson 1992 Singapore proc}, \cite{Standish 2000 private communication}, 
\cite{Pitjeva 2001 New value of delta-pi-dot}, but not by \cite{Krasinsky
1993 Motion of major planets} and \cite{Pitjeva1993 Mercury toppography}
(see table 1, section \ref{Tables})

But, once the quadrupolar correction is added, using (\ref
{perihelion_advance_parameter_i}), this leads to 
\begin{equation}
\Delta \omega _{GR}\in \left[ 43.00\emph{0}\ ;\ 43.01\emph{0}\right] \qquad 
\text{(arcsec/cy) for }J_{2}=(2.0\pm 0.4)\ 10^{-7}
\label{perihelion_advance_GR}
\end{equation}
which is now consistent with the observations given by \cite{Anderson 1992
Singapore proc}, \cite{Standish 2000 private communication}, \cite{Pitjeva
2001 New value of delta-pi-dot}, while still in agreement with \cite
{Krasinsky 1993 Motion of major planets} and \cite{Pitjeva1993 Mercury
toppography} (see table 1, section \ref{Tables}).\newline
This last result also shows that the theoretical prediction for $J_{2}$,
argued by the authors, is coherent with observations in the setting of
G.R.\medskip

\label{Important remark on the theoretical perihelion advance}An important
remark on values adopted for G.R.'s prediction of Mercury's perihelion
advance, $\Delta \omega _{0\ GR}$, in the past is given in \cite{Nobili 1986
Real value of Mercury perihelion advance}. The authors also interestingly
underline the following fact: \emph{``Although of theoretical interest, the
difference between these quoted predictions for Mercury's perihelion advance
has no observational consequence (for present methods of evaluation of
Mercury's perihelion shift)''. }Indeed, the predicted general relativistic
contribution to Mercury's perihelion advance is not an input in current
procedures testing gravitational theories with the dynamics of Mercury. In
modern ephemeris used to compute the motion of Mercury, the equations of
motion already include relativistic post-Newtonian terms which are
non-periodic. Those contribute to the secular variation of the orbital
elements, among which, the perihelion of Mercury. The post-Newtonian terms
in the ephemeris are modulated by a set of parameters (P.P.N. parameters
describing the gravitational theory, masses or initial conditions of the
planets, ...) that become part of a multiparameter least-squared fit to the
observational data (radar, optical data ...) in order to obtain an improved
determination of the parameters in the least-squares sense.\newline
However, it is impossible, presently, to fit simultaneously for both the
P.P.N.\ parameters and $J_{2}$, the two contributions, relativistic and
Newtonian respectively, to the perihelion shift being too correlated in the
case of Mercury alone (see section \ref{perihelion precession of minor
planets}). Thus, one can either directly test (fit) the P.P.N. parameters
assuming a given input value for $J_{2}$ in the ephemeris; or assume G.R. as
the gravitational theory and test $J_{2}$. In the last case, expression (\ref
{perihelion_advance}) together with (\ref{perihelion_advance_GR0}) are
useful to provide a value of $J_{2}$, once $\alpha =\beta =\gamma =1$ is
assumed.\newline
Nevertheless, the real general relativistic prediction for the perihelion
shift of Mercury ($\Delta \omega _{0\ GR}$) is given unequivocally by (\ref
{perihelion_advance_GR0}), according to present values of astrophysical
constants.

\subsubsection{Alternative theories to G.R. gravity.}

\qquad \label{Alternative theories to G.R, beta different from unity}\emph{%
General Relativity is often considered today as ``THE'' relativist theory of
gravitation.} This pure tensor theory corresponds to a Newtonian potential
that evolves as $1/r$, $r$ being the radial coordinate. The theory so far
agrees with all the observations made in our solar system. \emph{%
Nevertheless, the theory of General Relativity can not be the final theory
describing gravitation.}

First of all, \emph{from a theoretical point of view}, General Relativity
can not be quantified, and this makes it impossible to unify it with other
fundamental interactions.\newline
Moreover, the minimal choice of the Hilbert Einstein action, to which G.R.
corresponds, is not based upon any fundamental principle. Or to express it
in another way, it is evident that covariance and Newtonian fields
approximation alone do not determine uniquely the action. Equivalently,
nothing guaranties that the Newtonian potential is truly universal.\newline
Any other theory of gravitation would be valid too, as long as it would lead
to the same predictions as G.R. that have been tested in the solar system,
with maybe some departures from the Einsteinian theory on larger distance
scales.\newline
Also, let us warn that, from the formal point of view, the theory of General
Relativity is not invariant under conformal transformations. While, if we
wish to achieve the junction between particle physics, in which conformal
invariance plays a crucial role, and gravitation, we should consider a
theory of gravitation that incorporates this property.

\emph{From the experimental point of view}, let us notice that General
Relativity alone still can not reproduce the flat velocity distributions in
the vicinity of galaxies. The Newtonian potential would indeed predict a
decreasing distribution.\newline
We are thus confronted to the following dilemma: either we suppose the
existence of dark matter, either we modify the potential for galactic
distances. This second solution would immediately invalidate General
Relativity with a null cosmological constant.\newline
Let us also remark that the solution to the ``dark matter dilemma'' could
also be a combination of the two solutions cited here above.\medskip

\emph{In conclusion, according to the above arguments, it is fundamental to
conceive that alternative theories to General Relativity, that is to say }$%
\beta \neq 1$\emph{\ or/and }$\gamma \neq 1$\emph{, are truly not excluded
by the observations...} as the case of G.R., $\beta =\gamma =1$, is only a
particular spot in the allowed parameter space $\left( \beta ,\gamma
,J_{2}\right) $.\newline
This is illustrated by plotting ellipses representing the $1\sigma $, $%
2\sigma $ and $3\sigma $ confidence levels in the $\left( \beta ,\gamma
\right) $ plane, owing to Mercury's perihelion advance test, for a fixed
value of $J_{2}$ (See Figure 1 a, b, c). Further, adding constraints on $%
\gamma $ and $\beta $ coming from tests of the Nordtvedt effect and L.L.R.
data (see (\ref{estimated value of gamma from L.L.R. data}) and (\ref
{estimated value of beta from LLR and Nordtvedt effect})) allows to select a
portion of the ellipses in the $\left( \beta ,\gamma \right) $ plane.
However, G.R always belongs at least to the $3\sigma $ region in the allowed
parameter space $\left( \beta ,\gamma \right) $, according to the
theoretical bounds on $J_{2}$ adopted by the authors (see section \ref
{adopted theoretical value of J2}).\newline
Nevertheless, we can conclude that $\beta =1$ is not the only allowed
case.\medskip

\emph{Looking more in details at the contribution of }$\beta $\emph{\ to the
perihelion shift,} we see that the deviation, $1-\beta $, from G.R.'s value,
is, owing to the error bars, of the same order of magnitude as the
contribution of $J_{2}$.\newline
Thus two cases may be envisaged:\newline
Either $\beta <1$, which means that $\Delta \omega $ tends to be larger than 
$\Delta \omega _{GR}$, and the effect of $1-\beta $ adds to the contribution
of $J_{2}$.\newline
Either $\beta >1$, which means that $\Delta \omega $ tends to be smaller
than $\Delta \omega _{GR}$, and the effects of $1-\beta $ and $J_{2}$
substracts.

A possible consequence of $\beta $ being different from unity is the
Nordtvedt effect. Indeed, as soon as the combination of the
Eddington-Robertson parameters given by $\eta \equiv 4\beta -\gamma -3$ is
non null, the gravitational and inertial masses of a celestial body are no
longer the same (see \cite{Will 1993 Theory and experiment} and \cite{Will
2001 Theory and experiment update} and references there in).

New analysis of the Lunar Laser Ranging data (L.L.R.) by \cite{Williams 2001
Lunar Laser ranging} provides $\eta =+0.0002\pm 0.0009$, from which one may
deduce the acceptable range for $\beta $, using the value of $\gamma $ given
by (\ref{estimated value of gamma from L.L.R. data}): 
\begin{equation}
\beta \in \left[ 0.9993;1.0006\right] \text{\ .}
\label{estimated value of beta from LLR and Nordtvedt effect}
\end{equation}
This in turn allows us to infer a theoretical shift, $\Delta \omega $: 
\[
\Delta \omega \in \left[ 42.93\emph{2}\ ;\ 43.05\emph{7}\right] \qquad \text{%
(arcsec/cy) for }J_{2}=(2.0\pm 0.4)\ 10^{-7}\text{\ .} 
\]
We can see that it of course contains the particular case of G.R. ($\Delta
\omega _{GR}$), and that it is consistent with recent observations ($\Delta
\omega _{obs}$) (see table 1, section \ref{Tables}).

Alternatively, owing to the remark made in section \ref{Important remark on
the theoretical perihelion advance}, the fit of the most recent ephemeris
EPM2000 to accurate ranging observations concerning the motion of planets
(and in particular, the perihelion shift of Mercury), provide an
astonishingly precise estimation of the Eddington-Robertson parameters $%
\beta $ and $\gamma $ for a given theoretical value of $J_{2}$.\newline
Indeed, according to reference \cite{Pitjeva 2001 Modern Numerical
ephemerides}: 
\begin{equation}
\beta =1.0004\pm 0.0002\text{\ and }\gamma =1.0001\pm 0.0001\text{.}
\label{estimated value of beta and gamma from EPM2000}
\end{equation}
for a theoretical value of $2.0\ 10^{-7}$ for $J_{2}$, in agreement with (%
\ref{perihelion_advance_parameter_J2}). However, the uncertainties upon the
obtained parameters $\beta $ and $\gamma $ are formal deviations, and
realistic error bounds may be an order of magnitude larger. Moreover, this
estimation is rather tolerant regarding the assumed value of $J_{2}$.
Indeed, $\beta =1.000\pm 0.001$\ and $\gamma =1.0005\pm 0.0002$ have been
obtained using the test ephemeris which only differ from EPM2000 by the
solar oblateness $J_{2}=0.0$ \cite{Pitjeva 2001 New value of delta-pi-dot}!

\section{Inferring a dynamical value of the solar\newline
quadrupole moment in the setting of G.R.}

\qquad \label{inferring_J2}Conversely, one may think to infer the absolute
value of the quadrupole moment, $J_{2}$, which is necessary (owing the
allowed parameter space for $\beta $ and $\gamma $) to be in agreement with
observations. But, as mentioned in a remark in section \ref{Important remark
on the theoretical perihelion advance}, the purely relativistic contribution
($2\alpha ^{2}+2\alpha \gamma -\beta $) and the quadrupolar moment of the
Sun ($J_{2}$) are too correlated in the perihelion advance of Mercury to
lead simultaneously to interesting constraints on ($\beta $, $\gamma $) and $%
J_{2}$ separately.\newline
This is why, so far, a dynamical estimation of $J_{2}$ is made in the
setting of G.R.

In the particular case of G. R., the theory parametrized by $\alpha =\beta
=\gamma =1$, we find the results listed in table 3 (section \ref{Tables})
inferred from $\Delta \omega _{obs}$ (table 1, section \ref{Tables}) using
equation (\ref{perihelion_advance}).

Nevertheless, $J_{2}$ may not exceed the critical theoretical value of $3.0\
10^{-6}$ according to the argument given in \cite{Rozelot 1998 new results
on solar oblateness}\footnote{%
This estimation does not take into account a possible temporal dependence of 
$J_{2}$.\newline
If such a variability exists, the amplitude is, nevertheless, obviously
upper bounded by the critical value of $3.0\ 10^{-6}$.}, based upon the
accurate knowledge of the Moon's physical librations, for which the L.L.R.
data reaches accuracies at the milli-arcsecond level.\newline
Moreover, $J_{2}$ has to be positive to be in agrement with an oblate Sun.
So, only Standish and Pitjeva's last results, \cite{Pitjeva1993 Mercury
toppography}, \cite{Standish 2000 private communication} and \cite{Pitjeva
2001 Modern Numerical ephemerides}, give interesting dynamical constraints
upon $J_{2}$ (for the other authors, the error bars are too large or $J_{2}$
is negative, in contradiction with an oblate Sun). Namely: $J_{2}\leq 2.89\
10^{-7}$ for the EPM1988 ephemeris model, $J_{2}\leq 3.38\ 10^{-7}$ for the
DE200 model, $J_{2}=\left( 1.90\pm 0.16\right) \ 10^{-7}$ for the DE405
model and $J_{2}=\left( 2.453\pm 0.701\right) \ 10^{-7}$ for the more recent
EPM2000 model (see table 3, section \ref{Tables}). Those values are
compatible with the solar model dependent theoretical value of $J_{2}$, (\ref
{perihelion_advance_parameter_J2}), argued by the authors in section \ref
{adopted theoretical value of J2}.

\section{Increasing precision in the future}

\subsection{From Hipparcos to GAIA satellite: towards an astonishing
precision upon ($\gamma $, $\beta $,\ $J_{2}$)}

\subsubsection{Light deflection: $\gamma $}

\qquad Milli-arcsec astrometry is available since 1996 from Hipparcos
satellite data. The reduction of this data required the inclusion of stellar
aberration up to terms in $v/c^{2}$, as well as the correction (in $\frac{%
\alpha +\gamma }{2}$) due to the relativistic light deflection in the
gravitational field of the Earth and the Sun. Calculations for the Hipparcos
data were implicitly made in the setting of G.R. ($\alpha =\beta =\gamma =1$%
), thus allowing for this theory to be checked with a precision of $3\
10^{-3}$ on $\gamma $. This is of course less accurate than the results
based on V.L.B.I. measurements \cite{Robertson 1991 VLBI measurments of
gamma}, but Hipparcos opened the door to future micro-arcsec astrometry,
which can improve the precision upon $\gamma $ by several orders of
magnitude.\newline
Indeed, in the observational context of light deflection, the satellite GAIA 
\cite{GAIA 2000 March study report}, one cornerstone of ESA's Space Science
Programme, to be launched in 2009 (or at least no later than 2012) for a
five years mission, will increase the domain of observations by two orders
of magnitude in length (now, light deflection is tested on distances ranging
from $10^{9}$ to $10^{21}$ m) and six orders of magnitude in mass (now $1$
to $10^{13}\ M_{s}$). Moreover, GAIA, improving Hipparcos' performance, will
reduce the avoidance angle towards the Sun, thus allowing to measure
stronger light deflection effects with a reduced parallax correlation.%
\newline
This all results into an estimated accuracy of $5\ 10^{-7}$ on $\gamma $.

Notice that the quadrupolar moment of the Sun, $J_{2}$, has a contribution
to the light deflection that is negligeable in the case of GAIA, owing to
its non null avoidance angle\footnote{%
In the case of planets like Saturn or Jupiter being the deflector, the
contribution of $J_{2\ planet}$ to the light deflection effect is non
negligeable, due to the important magnitude of $J_{2\ planet}$ and to the
fact that grazing incidence is allowed.\medskip}.

\subsubsection{Perihelion precession for minor planets: $2\alpha
^{2}+2\alpha \gamma -\beta \ $and $J_{2}$}

\qquad \label{perihelion precession of minor planets}GAIA is also expected
to observe and discover several hundred thousand minor planets, mostly from
the Main Belt.\newline
All of them will acknowledge a perihelion shift (see equation (\ref
{perihelion_advance})), just like Mercury, but with a magnitude in respect
to the eccentricity, $e$, inclination, $i$, and semi-major axis, $a$, of
their own orbit.\newline
Thus, the relativistic correction \textit{per revolution} to the orbital
motion will only be significant for the Apollo, Aten and Armor groups, which
means of the same order of magnitude\footnote{%
For some exemples see \cite{GAIA 2000 March study report}, page 116 table
1.18.\medskip} as for Mercury. (In contrast, it will be about seven times
smaller for minor planets of the Main Belt). But unlike the Apollo and Aten
groups, the Armor group are not Earth-Orbit crossers.\newline
On the other side, \textit{the absolute precession rate} will be
approximately four times bigger for Mercury than for members of the Apollo
or Aten groups owing to their respective revolution periods (and more than
100 times bigger for the population of the Main Belt).\newline
Remark that the perihelion shift of the minor planet Icarus had already been
used in the past (as early as in 1968 \cite{Lieske 1969 Icarus and J2}) in
order to infer a dynamical value for $J_{2}$. But the non uniform
distribution of earlier observations\footnote{%
Based on photographic observations from 1949-1968, for \cite{Shapiro 1971 GR
and Icarus}, plus 7 Doppler-shift observations for reference \cite{Lieske
1969 Icarus and J2}; and additional observations during the encounter with
Earth in 1987 for reference \cite{Landgraf 1992 Estimation of J2 from Icarus}%
)} over the orbit of Icarus and Earth seriously affected the suitability of
(just) Icarus data in verifying G.R. or estimating $J_{2}$ independently
(see \cite{Shapiro 1965}, \cite{Shapiro 1968 Orbit of Icarus} and \cite
{Shapiro 1971 GR and Icarus}). So the estimations of $J_{2}$ were obtained
assuming G.R. (see table 4, section \ref{Tables}).

The advantages of measuring the perihelion shift of minor planets with GAIA,
in addition to Mercury's, are multiple.\newline
First, there will be, of course, an increased precision on individual
determinations of $\Delta \omega $, due to GAIA's technology but also to the
fact that minor planets are not as extended as Mercury, and so their
position can be measured more precisely.\newline
Secondly, a statistic on several tens of planets, which is a statistic on $%
\Delta \omega (a,\ e)$ (or $\delta (a,\ e)$), will allow to increase the
accuracy on the determination of $J_{2}$ and the combination ``$2\alpha
^{2}+2\alpha \gamma -\beta $'' separately. Remember that those two
contributions have different dependencies in ``$a\ \left( 1-e^{2}\right) $'' 
\cite{Gough 1982 Internal rotation and J2}.\newline
Thirdly, by studying the precession of the orbital plane of a minor planet
about the Sun's polar axis, due to the quadrupolar moment of the Sun ($J_{2}$%
) but unaffected by relativistic gravitation ($2\alpha ^{2}+2\alpha \gamma
-\beta $), one should be able to dynamically measure $J_{2}$ independently.
This effect being more easily discernible for moderately large values of the
inclination, $i$, minor planets like Icarus with a large value of $i$ ($%
i\simeq 16{{}^{\circ }}$) would be truly adequate (\cite{Dicke 1965 Icarus
and Relativity}, \cite{Shapiro 1965}, \cite{Shapiro 1968 Orbit of Icarus}).

A dedicated simulation still has to be performed to assess the real
capabilities of GAIA in that field. But, so far, an estimation of a
precision of $10^{-4}$ on the combination ``$2\alpha ^{2}+2\alpha \gamma
-\beta $'' from individual determinations of $\Delta \omega $, seems
reasonable; moreover, $10^{-5}$ should be attainable thanks to statistics on
several tens of planets.

From the point of view of $J_{2}$, GAIA should be more precise than $10^{-7}$%
, but the accuracy is difficult to assess without an extensive simulation on
the available sampling of ``$a\ \left( 1-e^{2}\right) $''. Nevertheless,
through measurement of perihelion advances, GAIA should provide a more
accurate dynamical and solar model independent determination of $J_{2}$, to
be confronted with solar model dependent predictions from, for example,
helioseismology data.

\subsubsection{Resulting constraint on $\beta $}

\qquad Using the independent constraint upon $\gamma $ obtained by GAIA from
light deflection, and the constraint upon ``$2\alpha ^{2}+2\alpha \gamma
-\beta $'' from perihelion shifts measured by GAIA, one should be able to
constraint $\beta $ with a precision of $3\ 10^{-4}-3\ 10^{-5}$. This is
about two orders of magnitude better than the present best determinations
due to L.L.R.(see equation (\ref{estimated value of beta from LLR and
Nordtvedt effect})) or direct fits of the data on the detection of $\eta $ 
\cite{Williams 2001 Lunar Laser ranging}.

\subsection{Alternative future direct measurements of $\gamma $}

\qquad GAIA will probably not deliver any results on $\gamma $ before the
end of its mission, but, in the mean time, other space or ground based
measurements like V.L.B.I.'s, will certainly improve the present
determination of $\gamma $. See table 5 (section \ref{Tables}) for proposed
space missions purely dedicated to the measurement of $\gamma $.

As an illustration of further prospects, we can cite the Astrodynamical
Space Test of Relativity using Optical Devices (ASTROD) \cite
{Bec-Borsenberger2000 ASTROD}, a proposal that has been submitted to ESA in
response to a ``Call for missions proposals for two Flexi-Missions'', but
which is not yet accepted. Such mission, using time-delay measurements
between two spacecrafts orbiting the Sun and the Earth, would certainly lead
to precisions of the order of $10^{-6}-10^{-7}$ on $\gamma $. But if the
stability of the clocks/lasers can be reduced to $10^{-18}$, then, using the
range data of ASTROD as an input for a better determination of the solar and
planetary parameters, one might dream to get a precision of $10^{-8}-10^{-9}$
on $\gamma $! Moreover, from the precise determination of orbits,
information could be given on the solar quadrupole moment, higher moments
and $\beta $. Again, providing ultra-stable clocks, precisions of the order
of $10^{-6}$ on $\beta $ and $4.5\ 10^{-8}$ on $J_{2}$ could be reached!

\subsection{A Mercury Orbiter mission: measuring $\beta $ independently from 
$\gamma $ as well as separating $\left( 2\alpha ^{2}+2\alpha \gamma -\beta
\right) $'s contribution from $J_{2}$'s}

\qquad \label{Mercury Orbiter}Scheduled to be launched in 2007 (or 2009) for
a 2-5 years mission, BepiColombo, has been accepted as E.S.A.'s Cornerstone
Mission \#5 in 1996 \cite{BepiColombo 2000 Study report}.\newline
It contains three spacecraft elements, among which a Mercury Planetary
Orbiter that will considerably help to reduce the error bars on the
Eddington-Robertson parameters $\beta $, $\gamma $ and the Sun's quadrupole
moment $J_{2}$.\newline
Indeed, an estimation of the accuracies attainable has been done thanks to a
full simulation of radio science experiments with calibration of solar
plasma noise, non gravitational accelerations and systematic effects.\newline
The measurement of $\beta $, $\gamma $, $J_{2}$ and $\eta $ is the output of
a complex orbit determination process in which radio-metric and calibration
acquired during the mission are used to provide a complete orbital solution
which includes the osculating orbital elements of the spacecraft and the
planet Mercury, as well as the harmonic coefficients of the planet's gravity
field (at least to the degree and order 25). Indeed, precision range and
range-rate measurements of BepiColombo will constrain the position of the
planet's center of mass with a precision of about $1$ m! Thus allowing a
truly precise knowledge of the orbital elements and the secular perihelion
shift of Mercury in particular.\newline
An additional advantage of a Mercury Orbiter, regarding the measure of the
perihelion shift, is that it would considerably reduce the time scale by
comparison to recent data observations that use time averages on a decade
time scale. This would permit an eventual observation of the variation of
the perihelion rotation due to a possible variation of the solar quadrupole
moment $J_{2}$.

From the view point of constraining the Eddington-Robertson parameter $%
\gamma $, time delay measurements of radio signals travelling from the
spacecraft to the Earth and back, combined with Doppler shift measurements
of photons should help get a yet more precise determination of $\gamma $.
BepiColombo should then be able to improve over Cassini's expected results
(see table 4, section \ref{Tables}), thanks to frequent solar conjunctions
during which the Doppler shift effect is maximum. A preliminary analysis of
the mission estimates that $\gamma $ could be accurately determined at $2.5\
10^{-6}$ \cite{BepiColombo 2000 Study report}.

The precise determination of Mercury's motion would also help measure the
Nordtvedt effect, $\eta $, with an expected accuracy of $2\ 10^{-5}$. This
combined with the values found for the perihelion advance ($2\alpha
^{2}+2\alpha \gamma -\beta $) and $\gamma $ would help lift the degeneracy
between $\beta $ and $J_{2}$.

Probing the gravitational field of Mercury at various distances from the
planet would also help separate the effects of $J_{2}$ from those of
relativistic gravitation ($2\alpha ^{2}+2\alpha \gamma -\beta $), owing to
their different dependency in the radial distance to Mercury \cite{Will 1993
Theory and experiment}.\newline
Following this idea, advantage can be taken of the large eccentricity of
Mercury's orbit to search for periodic orbital perturbations induced by $%
J_{2}$ and relativistic gravity \cite{Gough 1982 Internal rotation and J2}, 
\cite{Will 1993 Theory and experiment}.\newline
This is how $J_{2}$ would be determined independently by the BepiColombo
Mission: from the precise determination of the secular nodal precession of
the planet's orbital plane about the Sun's polar axis, due to the
quadrupolar moment of the Sun, but unaffected by purely relativistic
gravitation.\newline
All this should lead to a determination of $J_{2}$ with a precision of $2$ $%
10^{-9}$, (see \cite{Turyshev 1996 Mercury orbiter mission} and \cite
{BepiColombo 2000 Study report}).

Notice that the influence of Mercury's topography in determining the
perihelion precession has been stressed by some authors \cite{Pitjeva1993
Mercury toppography}, \cite{Pitjeva 2001 Modern Numerical ephemerides}. It
has to be taken into account when processing radar observations, as it might
help to reduce the systematic errors in ranging. So far, the scarcity of
radar observational data for Mercury restricts the accuracy of estimates for
the topographical contribution. But future Mercury orbiters, like
BepiColombo or N.A.S.A.'s discovery mission named Messenger\footnote{%
But, unlike BeppiColombo, it will not test G.R. nor measure the perihelion
precession... (see http:\TEXTsymbol{\backslash}\TEXTsymbol{\backslash}%
discovery.nasa.gov\TEXTsymbol{\backslash}messenger.html and section 9.4 in
reference \cite{BepiColombo 2000 Study report}).\medskip} (to be launched in
2004), could remedy to that problem by providing useful complementary data
upon the topography of the planet.

Finally, we shall cite an interesting proposal from article\textbf{\ }\cite
{Turyshev 1996 Mercury orbiter mission} (page 24, equation 40), that
suggests a measure of $\beta $\ independent of $\gamma $, testing the strong
equivalence principle, using a Mercury orbiter on a particular resonant
orbit.

\subsection{Satellites dedicated to $J_{2}$}

\qquad \label{Picard Satellite}Future solar probes are expected to determine
a more precise value for $J_{2}$. \newline
Indeed, the quadrupole moment of the Sun can be measured dynamically by
sending and accurately tracking a probe, equipped with a drag-free guidance
system, to within a few solar radii of the solar center. $J_{2}$ is then
inferred from the precise determination of the trajectory.\newline
Alternatively, $J_{2}$ can be inferred from in orbit measurements of solar
properties. But in this case, the reduction of such a measurement will
require a better understanding on how solar density models and rotational
laws influence the multipole expansion of the external gravitational field 
\cite{Ulrich 1981 Solar gravitational figure}.\newline
For example, the micro-satellite Picard is a C.N.E.S.\footnote{%
Centre National d'Etudes Spatiales (France)\medskip} mission\textbf{,} due
for flight by the end of 2005. The expected mission lifetime is 3 to 4 years
with a possible extension to 6 years.\newline
The aim of Picard \cite{Dame 2001 Picard}, is to perform in orbit
simultaneous accurate and absolute measurements of the solar diameter,
differential rotation and irradiance, in addition to low frequency
helioseismology, as a permanent viewing of the Sun from a G.T.O. orbit
should allow the detection of g-modes.\newline
Picard should be able to measure $J_{2}$ with of precision of $10^{-8}$.%
\newline
Notice also that the diameter measurements will be obtained at any latitude
(sunspots and faculae at limb removed) which should allow the detection of a
latitudinal variation of the diameter and thus, the quadrupole moment, as
predicted by the theoretical model used by the authors \cite{Godier-Rozelot
1999b Relationship J2 and layers of the Sun}.\newline
Moreover, Picard will observe the Sun in different wavelength bands, among
which the band used for measurements of the Sun's diameter from the ground, $%
535.7$ nm. This will permit one to compare space-measurements with
ground-based ones in order to correct for atmospheric perturbations, and so,
to eventually re-calibrate the existing data over the former solar cycle
dependencies.\newline
Thus, the Picard mission is clearly designed to solve some of the problems
mentioned in section \ref{controversies on J2 determination} for the
determination of $J_{2}$.

\section{CONCLUSIONS}

\qquad We have seen in this article that, \textbf{so far, the theory of
General Relativity is not excluded by observations}. However G.R. only
represents one possible point in the still large allowed parameter space $%
(\beta $, $\gamma $, $J_{2})$ and alternative theories are also permitted.%
\newline
More precisely:

Future space experiments cited in the last section of this article will
considerably reduce the parameter space in the near future, and so impose an
even more rigorous test to G.R..

As a determination of the solar quadrupole moment is concerned, \textbf{we
have stressed the importance of confronting a dynamical determination of }$%
\mathbf{J}_{\mathbf{2}}$\textbf{, independent of the solar model (obtained
from perihelion advances or motion of spacecrafts), to other solar model
dependent values}.\newline
Presently, this dynamical determination of $J_{2}$ is still dependent upon
the gravitational theory ($2\alpha ^{2}+2\alpha \gamma -\beta $), but the 
\textbf{future GAIA or BepiColombo missions should be able to separate those
two effects, and so obtain a determination of }$\mathbf{J}_{\mathbf{2}}$%
\textbf{\ that is independent of the gravitational theory}.\newline
So far too, the authors can say that their estimated theoretical value of $%
J2=\left( 2.0\pm 0.4\right) \ 10^{-7}$, which is solar model dependent,
together with the estimated error bars, is completely coherent with the
estimated dynamical value resulting from Mercury perihelion advance in the
setting of General Relativity.

We conclude by reminding that, if future observations confirm the time
dependency of $J2$ (for example, a periodicity with the sunspots cycle),
this effect will have to be taken into account when using perihelion advance
as a test for G.R. (or alternatively, to estimate $J2$). This so far has not
been the case, as dynamically estimated values of $J2$ come from a mean over
several decades (see for example \cite{Landgraf 1992 Estimation of J2 from
Icarus}).

\section{ACKNOWLEDGMENTS}

\qquad The authors thank M. Standish for comments on the determination of
the perihelion advance from radar data and for providing his determination
of Mercury's perihelion advance; E. Dipietro for usefull discussions.They
are also grateful to E.V. Pitjeva for precisions given about her articles 
\cite{Pitjeva 2001 New value of delta-pi-dot} and for providing her new
value of $\Delta \stackrel{\bullet }{\pi }$, yet to be published \cite
{Pitjeva 2001 Modern Numerical ephemerides}.

This work was done under a I.I.S.N. research assistantship\ for one of us
(S. Pireaux).

\section{TABLES AND FIGURES}

\label{Tables}

\[
\begin{tabular}{c}
\hline
\begin{tabular}{c}
- Table 1 - \\ 
Inferred correction to G.R.'s prediction for Mercury's perihelion advance
\end{tabular}
\\ \hline
\begin{tabular}[t]{l|l|l|l}
References & 
\begin{tabular}[t]{l}
Perihelion \\ 
Advance \\ 
$\Delta \omega _{obs}$ \\ 
{\small (arcsec/cy)}
\end{tabular}
& 
\begin{tabular}[t]{l}
\\ 
\\ 
$\Delta \omega _{obs}-\Delta \omega _{0\ GR}$ \\ 
{\small (arcsec/cy)}
\end{tabular}
& 
\begin{tabular}[t]{l}
$\delta $ \\ 
$\shortparallel $ \\ 
$\Delta \omega _{obs}/\Delta \omega _{0\ GR}$%
\end{tabular}
\\ \hline
& \multicolumn{1}{|c|}{} & \multicolumn{1}{|c|}{} & \multicolumn{1}{|c}{} \\ 
\begin{tabular}{l}
{\small \cite{Newcomb 1895-1898 Perihelion advance and J2}}
\end{tabular}
& \multicolumn{1}{|c|}{$\sim ${\small 43.37}} & \multicolumn{1}{|c|}{$\sim $%
{\small 0.39}} & \multicolumn{1}{|c}{$\sim ${\small 1.01}} \\ 
\begin{tabular}{l}
{\small \cite{Clemence 1943 Perihelion advance}}
\end{tabular}
& \multicolumn{1}{|c|}{{\small 42.84}$\pm ${\small 1.01}} & 
\multicolumn{1}{|c|}{{\small -0.14}$\pm ${\small 1.01}} & 
\multicolumn{1}{|c}{{\small 0.997}$\pm ${\small 0.023}} \\ 
\begin{tabular}{l}
{\small \cite{Clemence 1947 Perihelion advance}}
\end{tabular}
& \multicolumn{1}{|c|}{{\small 42.57}$\pm ${\small 0.96}} & 
\multicolumn{1}{|c|}{{\small -0.41}$\pm ${\small 0.96}} & 
\multicolumn{1}{|c}{{\small 0.990}$\pm ${\small 0.022}} \\ 
\begin{tabular}{l}
{\small \cite{Duncombe 1958 Mercury (Planet)}}
\end{tabular}
& \multicolumn{1}{|c|}{{\small 43.10}$\pm ${\small 0.44}} & 
\multicolumn{1}{|c|}{{\small +0.12}$\pm ${\small 0.44}} & 
\multicolumn{1}{|c}{{\small 1.003}$\pm ${\small 0.010}} \\ 
\begin{tabular}{l}
{\small \cite{Wayman 1966 Determination of the Inertial frame of reference}}
\end{tabular}
& \multicolumn{1}{|c|}{{\small 43.95}$\pm ${\small 0.41}} & 
\multicolumn{1}{|c|}{{\small +0.97}$\pm ${\small 0.41}} & 
\multicolumn{1}{|c}{{\small 1.023}$\pm ${\small 0.010}} \\ 
\begin{tabular}{l}
{\small \cite{Shapiro 1972 Mercury perihelion advance and radar data}}
\end{tabular}
& \multicolumn{1}{|c|}{{\small 43.15}$\pm ${\small 0.30}} & 
\multicolumn{1}{|c|}{{\small +0.17}$\pm ${\small 0.30}} & 
\multicolumn{1}{|c}{{\small 1.004}$\pm ${\small 0.007}} \\ 
\begin{tabular}{l}
{\small \cite{Morrison 1975 Analysis of the transits of Mercury}}
\end{tabular}
& \multicolumn{1}{|c|}{{\small 41.90}$\pm ${\small 0.50}} & 
\multicolumn{1}{|c|}{{\small -1.08}$\pm ${\small 0.50}} & 
\multicolumn{1}{|c}{{\small 0.975}$\pm ${\small 0.016}} \\ 
\begin{tabular}{l}
{\small \cite{Shapiro 1976 equivalence principle}}
\end{tabular}
& \multicolumn{1}{|c|}{{\small 43.11}$\pm ${\small 0.21}} & 
\multicolumn{1}{|c|}{{\small +0.13}$\pm ${\small 0.21}} & 
\multicolumn{1}{|c}{{\small 1.003}$\pm ${\small 0.005}} \\ 
\begin{tabular}{l}
{\small \cite{Anderson 1978 Tests of GR using astrometric and radiometric
observations}}
\end{tabular}
& \multicolumn{1}{|c|}{{\small 43.3}$\pm ${\small 0.2}} & 
\multicolumn{1}{|c|}{{\small +0.32}$\pm ${\small 0.2}} & \multicolumn{1}{|c}{%
{\small 1.007}$\pm ${\small 0.005}} \\ 
$\left. 
\begin{array}{l}
\text{{\small \cite{Bretagnon 1982a Perihelion advance}}} \\ 
\text{{\small \cite{Narlikar 1985 N-body calculation}}}
\end{array}
\right\} $ & \multicolumn{1}{|c|}{{\small 45.40}$\pm ${\small 0.05}} & 
\multicolumn{1}{|c|}{{\small +2.42}$\pm ${\small 0.16}} & 
\multicolumn{1}{|c}{{\small 1.056}$\pm ${\small 0.001}} \\ 
$\left. 
\begin{array}{l}
\text{{\small \cite{Bretagnon 1982b Perihelion advance}}} \\ 
\text{{\small \cite{Rana 1987 motion of the node}}}
\end{array}
\right\} $ & \multicolumn{1}{|c|}{{\small 45.25}$\pm ${\small 0.05}} & 
\multicolumn{1}{|c|}{{\small +2.27}$\pm ${\small 0.05}} & 
\multicolumn{1}{|c}{{\small 1.053}$\pm ${\small 0.002}} \\ 
\begin{tabular}{l}
{\small \cite{Krasinsky 1986 Relativistic effects on planetary observations}}
\end{tabular}
& \multicolumn{1}{|c|}{} & \multicolumn{1}{|c|}{} & \multicolumn{1}{|c}{} \\ 
$\quad 
\begin{array}[t]{l}
\text{{\small EPM1988}} \\ 
\text{{\small DE200}}
\end{array}
$ & \multicolumn{1}{|c|}{$
\begin{array}[t]{l}
\text{{\small 42.83}}\pm \text{{\small 0.12}} \\ 
\text{{\small 42.77}}\pm \text{{\small 0.12}}
\end{array}
$} & \multicolumn{1}{|c|}{$
\begin{array}[t]{l}
\text{{\small -0.15}}\pm \text{{\small 0.12}} \\ 
\text{{\small -0.21}}\pm \text{{\small 0.12}}
\end{array}
$} & \multicolumn{1}{|c}{$
\begin{array}[t]{l}
\text{{\small 0.997}}\pm \text{{\small 0.003}} \\ 
\text{{\small 0.995}}\pm \text{{\small 0.003}}
\end{array}
$} \\ 
\begin{tabular}{l}
{\small \cite{Rana 1987 motion of the node}}
\end{tabular}
& \multicolumn{1}{|c|}{{\small 45.47}$\pm ${\small 0.09}} & 
\multicolumn{1}{|c|}{{\small +2.49}$\pm ${\small 0.09}} & 
\multicolumn{1}{|c}{{\small 1.058}$\pm ${\small 0.002}} \\ 
\begin{tabular}{l}
{\small \cite{Anderson 1987 Ephemeris Mercury}}
\end{tabular}
& \multicolumn{1}{|c|}{{\small 42.92}$\pm ${\small 0.20}} & 
\multicolumn{1}{|c|}{{\small -0.06}$\pm ${\small 0.20}} & 
\multicolumn{1}{|c}{{\small 0.999}$\pm ${\small 0.005}} \\ 
\begin{tabular}{l}
{\small \cite{And 1991 IAU proc}}
\end{tabular}
& \multicolumn{1}{|c|}{{\small 42.94}$\pm ${\small 0.20}} & 
\multicolumn{1}{|c|}{{\small -0.04}$\pm ${\small 0.20}} & 
\multicolumn{1}{|c}{{\small 0.999}$\pm ${\small 0.005}} \\ 
\begin{tabular}{l}
{\small \cite{Anderson 1992 Singapore proc}}
\end{tabular}
& \multicolumn{1}{|c|}{{\small 43.13}$\pm ${\small 0.14}} & 
\multicolumn{1}{|c|}{{\small +0.15}$\pm ${\small 0.14}} & 
\multicolumn{1}{|c}{{\small 1.003}$\pm ${\small 0.003}} \\ 
\begin{tabular}{l}
{\small \cite{Krasinsky 1993 Motion of major planets}}
\end{tabular}
$:$ & \multicolumn{1}{|c|}{} & \multicolumn{1}{|c|}{} & \multicolumn{1}{|c}{}
\\ 
$\quad 
\begin{array}[t]{l}
\text{{\small EPM1988}} \\ 
\text{{\small DE200}}
\end{array}
$ & \multicolumn{1}{|c|}{$
\begin{array}[t]{l}
\text{{\small 42.985}}\pm \text{{\small 0.061}} \\ 
\text{{\small 42.978}}\pm \text{{\small 0.061}}
\end{array}
$} & \multicolumn{1}{|c|}{$
\begin{array}[t]{l}
\text{{\small +0.004}}\pm \text{{\small 0.061}} \\ 
\text{{\small -0.003}}\pm \text{{\small 0.061}}
\end{array}
$} & \multicolumn{1}{|c}{$
\begin{array}[t]{l}
\text{{\small 1.0001}}\pm \text{{\small 0.0014}} \\ 
\text{{\small 0.9999}}\pm \text{{\small 0.0014}}
\end{array}
$} \\ 
\begin{tabular}{l}
{\small \cite{Pitjeva1993 Mercury toppography}}
\end{tabular}
{\small :} & \multicolumn{1}{|c|}{} & \multicolumn{1}{|c|}{} & 
\multicolumn{1}{|c}{} \\ 
$\quad 
\begin{array}[t]{l}
\text{{\small EPM1988}} \\ 
\text{{\small DE200}}
\end{array}
$ & \multicolumn{1}{|c|}{$
\begin{array}[t]{l}
\text{{\small 42.964}}\pm \text{{\small 0.052}} \\ 
\text{{\small 42.970}}\pm \text{{\small 0.052}}
\end{array}
$} & \multicolumn{1}{|c|}{$
\begin{array}[t]{l}
\text{{\small -0.017}}\pm \text{{\small 0.052}} \\ 
\text{{\small -0.011}}\pm \text{{\small 0.052}}
\end{array}
$} & \multicolumn{1}{|c}{$
\begin{array}[t]{l}
\text{{\small 0.9996}}\pm \text{{\small 0.0012}} \\ 
\text{{\small 0.9997}}\pm \text{{\small 0.0012}}
\end{array}
$} \\ 
\begin{tabular}{l}
{\small \cite{Standish 2000 private communication}}
\end{tabular}
& \multicolumn{1}{|c|}{} & \multicolumn{1}{|c|}{} & \multicolumn{1}{|c}{} \\ 
$\quad 
\begin{array}{l}
\text{{\small DE405}}
\end{array}
$ & \multicolumn{1}{|c|}{{\small 43.004}$\pm ${\small 0.002}} & 
\multicolumn{1}{|c|}{{\small +0.023}$\pm ${\small 0.002}} & 
\multicolumn{1}{|c}{{\small 1.00054}$\pm ${\small 0.00005}} \\ 
\begin{tabular}{l}
{\small \cite{Pitjeva 2001 New value of delta-pi-dot}}
\end{tabular}
{\small :} & \multicolumn{1}{|c|}{} & \multicolumn{1}{|c|}{} & 
\multicolumn{1}{|c}{} \\ 
$\quad 
\begin{array}{l}
\text{{\small EPM2000}}
\end{array}
$ & \multicolumn{1}{|c|}{{\small 43.0115}$\pm ${\small 0.0085}} & 
\multicolumn{1}{|c|}{{\small +0.0305}$\pm ${\small 0.0085}} & 
\multicolumn{1}{|c}{{\small 1.00071}$\pm ${\small 0.00020}}
\end{tabular}
\\ \hline
\end{tabular}
\]

\begin{quote}
Table 1. {\small In the past, planetary motions, necessary to infer the
value of }$\Delta \omega _{obs}${\small , were modeled with classical
analytical theories. Presently, more precise space experiments require
accurate numerical ephemeris. Those are made possible thanks to new
astrometric methods (radar ranging, Lunar Laser Ranging, V.L.B.I..
measurements) that led to ranging data uncertainties of only a few meters.}%
\newline
{\small An interesting review of EPM and DE numerical ephemeris can be found
in reference \cite{Pitjeva 2001 Modern Numerical ephemerides}. It discusses
the history of planetary motion modeling and describes which data (optical,
radar, L.L.R. or from space-craft) has been used for the different ephemeris.%
}

{\small Following this historical path, references \cite{Newcomb 1895-1898
Perihelion advance and J2}, \cite{Clemence 1943 Perihelion advance}
(calculated for Julian year J1900.00), \cite{Clemence 1947 Perihelion
advance} (J1850.00) and \cite{Bretagnon 1982a Perihelion advance} (J1900.00)
used an analysis of the observed data on Mercury which was biased by the
assumed theory of gravitation.\newline
In \cite{Clemence 1943 Perihelion advance}, \cite{Duncombe 1958 Mercury
(Planet)}, \cite{Wayman 1966 Determination of the Inertial frame of
reference} and \cite{Morrison 1975 Analysis of the transits of Mercury}, the
same old analytical perturbation theory as Newcomb's, \cite{Newcomb
1895-1898 Perihelion advance and J2}, was used. Or, in \cite{Clemence 1947
Perihelion advance}, the results are based on Doolittle's calculations of
the Newtonian motion, with certain corrections \cite{Doolittle 1925 Secular
elements}. Those methods were not as adequate as present numerical
computations. While the semi-analytical theory developed in \cite{Lestrade
1982 Perihelion advance} did not use sufficiently accurate equations of
motions of planets.\newline
See arguments given in \cite{Narlikar 1985 N-body calculation} and \cite
{Rana 1987 motion of the node}.\medskip }

\qquad {\small To each reference in table 1 corresponds a value of the
advance of Mercury's perihelion deduced from observational data: }$\Delta
\omega _{obs}${\small .\newline
For each of them, one may compute the possible corrective factor, }$\delta $%
{\small , to the prediction due to Einstein's Gravity (G.R.), }$\Delta
\omega _{0\ GR}=42.98\emph{1}${\small \ arcsec/cy, which does not include
the quadrupolar (}$J_{2}${\small ) correction.\newline
Notice that light deflection measurements constraint }$\gamma ${\small \ to }%
$0.9996\pm 0.0017${\small \ \cite{Lebach 1995 Light deflection}; while
theoretical predictions for the solar quadrupole moment, taking into account
surfacic and internal differential rotation, give }$J_{2}=\left( 2.0\pm
0.4\right) \ 10^{-7}${\small , which means that the solar correction to the
perihelion advance, -}$\frac{R_{s}^{2}}{R\ \alpha \ a\left( 1-e^{2}\right) }%
\ J_{2}\ \left( 3\sin ^{2}i-1\right) =2.821\emph{8}\ 10^{-4}\frac{J_{2}}{%
10^{-7}}${\small \ is }$2.821\emph{8}\ 10^{-4}\ \left( 2.0\pm 0.4\right) $%
{\small .\newline
New L.L.R. data \cite{Williams 2001 Lunar Laser ranging}, on their side,
provide}\newline
$\beta \in \left[ 0.9993;1.0006\right] ${\small .\medskip }

{\small Here follow some comments on how the values, }$\Delta \omega _{obs}$%
{\small , given in table 1, are inferred from the given references:}

{\small In \cite{Clemence 1943 Perihelion advance}, the discordance O-C
between observations and the modeling theory used by Clemence is }$-0".07\pm
0.41${\small . In this theory, Clemence took }$42".91${\small \ for the
general relativistic contribution to the precession. }$\Delta \omega _{obs}$%
{\small \ is thus the sum of those two numbers, where Clemence's estimation
of the probable error in the uncertainty in the masses, }$\pm 0.6${\small ,
is taken into account.}

{\small The values given in this table for \cite{Clemence 1947 Perihelion
advance} result from Clemence's value $42".56\pm 0.94$\ (which is the
difference between the total perihelion precession observed, $5599".74\pm
0.41$, and the Newtonian contribution of the other Planets plus the effect
of the solar oblateness, $5557".18\pm 0.85$) from which Clemence's erroneous
estimation of $J_{2}$'s contribution, $0".010\pm 0.02$, has been removed.}

{\small As reference \cite{Wayman 1966 Determination of the Inertial frame
of reference} is concerned, }$\Delta \omega _{obs}$\ {\small is obtained by
substracting the total Newtonian contribution of planets, }$531".26${\small %
\ (that had been recalculated by the author of \cite{Wayman 1966
Determination of the Inertial frame of reference} using Marsden's masses for
planets) and the precession of the equinoxes as calculated by Newcomb \cite
{Newcomb 1895-1898 Perihelion advance and J2}, }$5024".53${\small , from the
total precession observed, }$5599".74${\small , as cited from \cite{Clemence
1943 Perihelion advance}.}

{\small The values given for \cite{Shapiro 1972 Mercury perihelion advance
and radar data} are obtained from their given estimation of $\delta $\
(written in their article as $\lambda _{p}$ with }$J_{2}=0${\small ) that
includes a correction for a typical representation of the topography of the
Planet Mercury, using equations (\ref{perihelion_advance}) and (\ref
{perihelion_advance_GR0}).}

{\small In references \cite{Bretagnon 1982a Perihelion advance} and \cite
{Bretagnon 1982b Perihelion advance}, the perihelion advance had been
recalculated at J1900.00, using contemporary values for planetary masses
(see \cite{Narlikar 1985 N-body calculation} and \cite{Rana 1987 motion of
the node} respectively). For them, the constant rate of the perihelion
advance due to the equinoxes precession is taken to be respectively $%
5029^{"}.0966+2.2223T$ (cited in \cite{Bretagnon 1982a Perihelion advance}
from \cite{Lieske 1977 Expression for precession quantities}) and $%
5029^{"}.0966+2.2274T$ (cited in \cite{Bretagnon 1982b Perihelion advance}
from \cite{Bretagnon 1981 Calcul de la precession}), where $T$\ is in Julian
centuries with reference to the initial epoch J2000; and the N-body
Newtonian contribution from planets, $528^{"}.95$, from \cite{Narlikar 1985
N-body calculation}.}

{\small Notice that in the following articles, \cite{Krasinsky 1986
Relativistic effects on planetary observations}, \cite{Krasinsky 1993 Motion
of major planets}, \cite{Pitjeva1993 Mercury toppography}, the authors
improperly writes $42.95$ (arcsec/cy) for $\Delta \omega _{0\ GR}$. \newline
But the values for $\Delta \stackrel{\bullet }{\pi }$ given in those
articles are not affected (see remark in section \ref{Important remark on
the theoretical perihelion advance})$.$\ Indeed, the correction to the
perihelion motion, written as $\Delta \stackrel{\bullet }{\pi }$ in those
articles, has been obtained by fitting the EPM88 and DE200 numerical
ephemerides to all radar observational data (from 1964 to 1989 for
Pitjeva's) with the needed parameters (the elements of planetary orbits, the
radii of planets, the value of AU, etc.). The correction to the perihelion
motion is thus the ``observed'' deviation from the value of Mercury
perihelion advanced obtained from EPM88 ephemerides with zero $J_{2}$\
assumed on the time interval mentioned. In this present table, we computed $%
\Delta \omega _{obs}$\ as $\Delta \omega _{EPM1988}+$$\Delta \stackrel{%
\bullet }{\pi }$, where is $\Delta \omega _{EPM1988}$\ the mean value of the
perihelion advance obtained from EPM1988 ephemerides, namely, $%
42".9806\simeq 42".981$\ (arcsec/cy).\newline
As far as the estimation of the quadrupole moment, $J_{2}$, is concerned, it
has been incorrectly done, in those articles, with the value $42".95$
(arcsec/cy). But it can be recomputed afterwards; assuming G.R. (fixed value
for $\alpha =\beta =\gamma =1$) and using the differentiation of formula (%
\ref{perihelion_advance}) with the proper value for $\Delta \omega _{0\ GR}$
given in (\ref{perihelion_advance_GR0}):\newline
$\Delta J_{2}=\frac{\Delta \stackrel{\bullet }{\pi }}{\Delta \omega _{0\
GR}\ 2.821\emph{8}}\ 10^{-3}$ and $J_{2}=J_{2\ EPM1988}+\Delta J_{2}$.%
\newline
The rounded up result that appears in those articles are not much affected
(see our table 3).}

{\small In \cite{Pitjeva 2001 Modern Numerical ephemerides}, the new
estimation of $\Delta \stackrel{\bullet }{\pi }=+0".0055\pm 0".0085$ is
obtained from a fit of EPM2000 numerical ephemeris to radar observational
data over the period 1961-1997. But now, $J_{2\ EPM2000}=2.0\ 10^{-7}$\ was
assumed and thus, the mean value of the perihelion shift, $\Delta \omega
_{EPM2000}\simeq $$43".0060$, is different from $\Delta \omega _{0\ GR}$.}

{\small In \cite{Standish 2000 private communication}, a value of $J_{2\
DE405}=2.0\ 10^{-7}$ was assumed. But the estimated perihelion advance given
by M. Standish, $42".980\pm 0".002$, contains solely the purely relativistic
contribution. The quadrupole moment's, $J_{2\ DE405}$'s, contribution must
thus be added.}
\end{quote}

\pagebreak

\begin{landscape}

\begin{tabular}{c}
\begin{tabular}{lll}
\FRAME{itbpF}{2.5746in}{2.6377in}{0in}{}{}{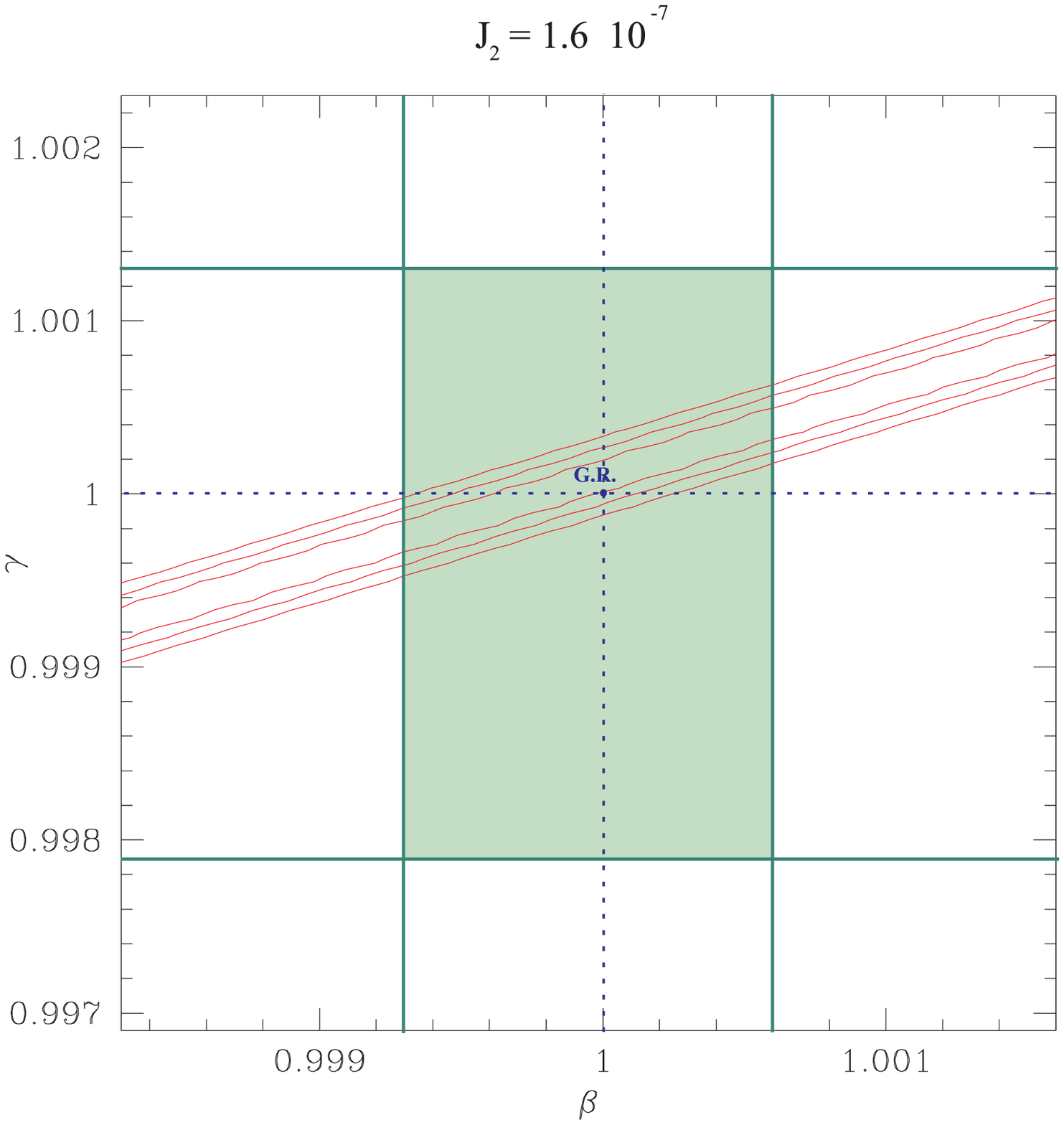}{%
\special{language "Scientific Word";type "GRAPHIC";maintain-aspect-ratio
TRUE;display "ICON";valid_file "F";width 2.5746in;height 2.6377in;depth
0in;original-width 0pt;original-height 0pt;cropleft "0";croptop
"1";cropright "1";cropbottom "0";filename
'figure1.eps';file-properties "XNPEU";}} & \FRAME{itbpF%
}{2.5746in}{2.6377in}{-0.0173in}{}{}{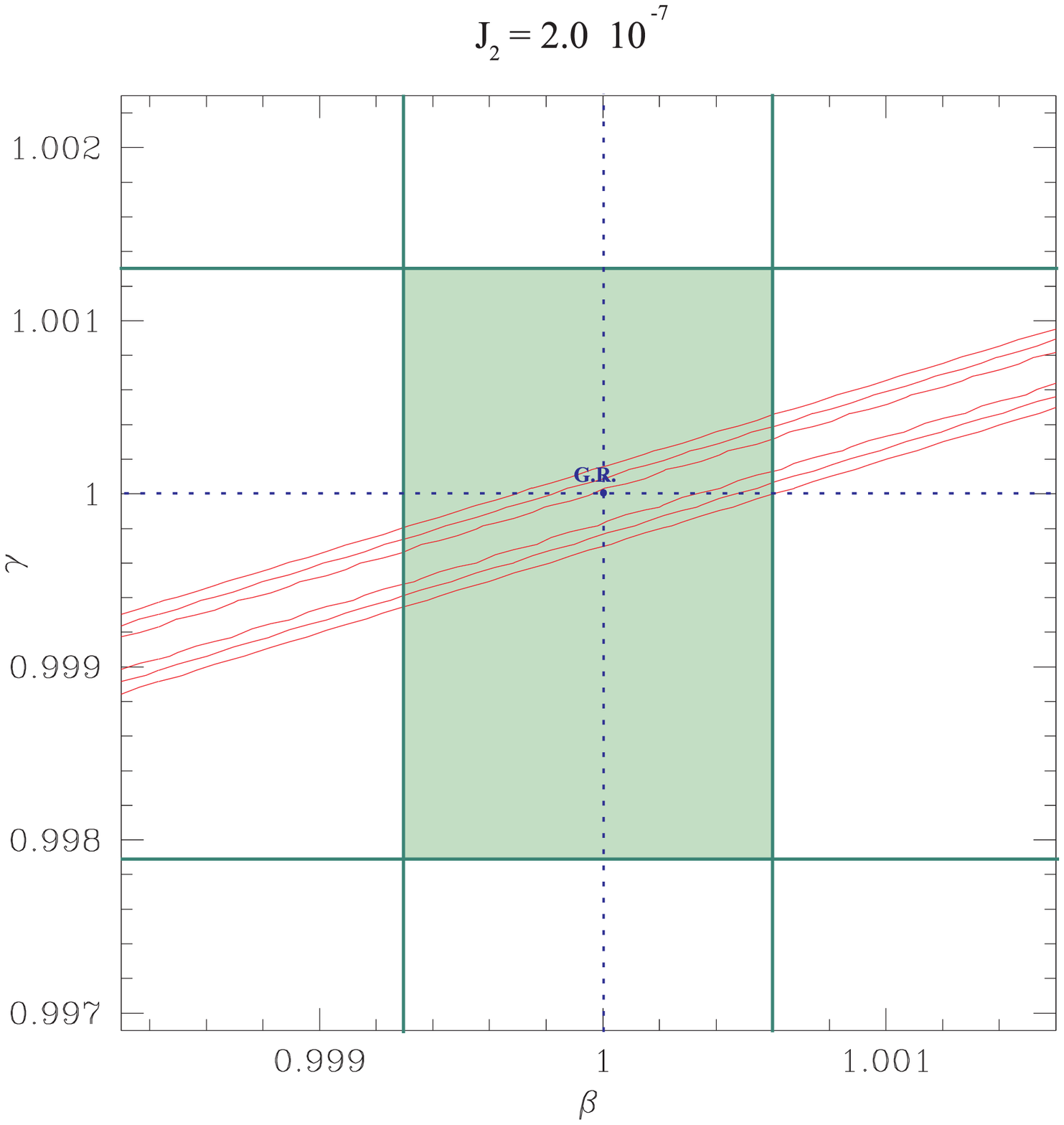}{\special{language
"Scientific Word";type "GRAPHIC";maintain-aspect-ratio TRUE;display
"ICON";valid_file "F";width 2.5746in;height 2.6377in;depth
-0.0173in;original-width 0pt;original-height 0pt;cropleft "0";croptop
"1";cropright "1";cropbottom "0";filename
'figure2.eps';file-properties "XNPEU";}} & \FRAME{%
itbpF}{2.5746in}{2.6377in}{0in}{}{}{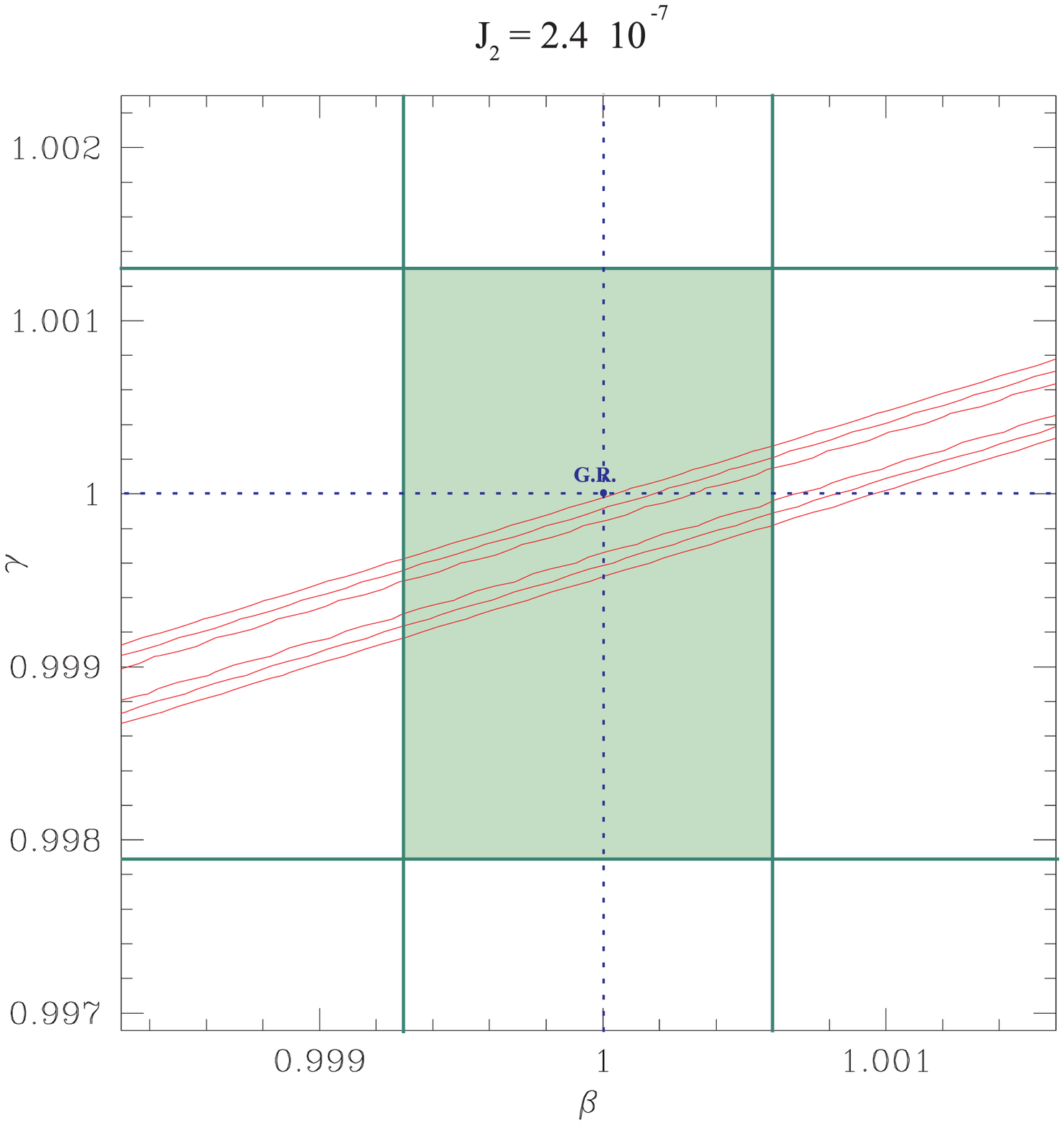}{\special{language
"Scientific Word";type "GRAPHIC";maintain-aspect-ratio TRUE;display
"ICON";valid_file "F";width 2.5746in;height 2.6377in;depth
0in;original-width 0pt;original-height 0pt;cropleft "0";croptop
"1";cropright "1";cropbottom "0";filename
'figure3.eps';file-properties "XNPEU";}}
\end{tabular}
\\ 
\begin{tabular}[t]{l}
Figure 1. {\footnotesize For a given value of }$J_{2}${\footnotesize , the
perihelion advance of Mercury constitutes a test of the P.P.N. parameters }$%
\beta ${\footnotesize \ and }$\gamma ${\footnotesize .} \\ 
{\footnotesize In the }$\beta ${\footnotesize \ and }$\gamma ${\footnotesize %
\ plane (}$\alpha ${\footnotesize \ set to 1), we have plotted 1}$\sigma $%
{\footnotesize \ (the smallest), 2}$\sigma ${\footnotesize \ and 3}$\sigma $%
{\footnotesize \ (the largest) confidence level ellipses.} \\ 
{\footnotesize Those are based on the values for Mercury's observed
perihelion advance, }$\Delta \omega _{obs}${\footnotesize , given in table
1. Notice however that} \\ 
{\footnotesize the value given by \cite{Newcomb 1895-1898 Perihelion advance
and J2} as well as those given by }$\text{{\small \cite{Bretagnon 1982a
Perihelion advance}}}${\footnotesize , }$\text{{\small \cite{Bretagnon 1982b
Perihelion advance} }}${\footnotesize and \cite{Rana 1987 motion of the node}
have not been taken into} \\ 
{\footnotesize account. Indeed, the first cited reference did not contain
any error bars estimation; the other ones used an improper method} \\ 
{\footnotesize to evaluate }$\Delta \omega _{obs}${\footnotesize \ (see
comments of table 1) and the error bars they provide are truely not
realistic ones. Remark also that} \\ 
{\footnotesize the position of the ellipses varies according to the value of 
}$J_{2}${\footnotesize \ chosen; but, their orientation is determined by the
combination} \\ 
{\footnotesize (}$2\alpha ^{2}+2\alpha \gamma -\beta ${\footnotesize ) that
appears in the expression for }$\Delta \omega ${\footnotesize .} \\ 
{\footnotesize Nevertheless, G.R. is still in the 3}$\sigma ${\footnotesize %
\ contours for the allowed theoretical values of }$J_{2}${\footnotesize \
argued by the authors (see section \ref{adopted theoretical value of J2}).}
\\ 
{\footnotesize Fig. 1 a, b, c represent the confidence contours for }$\beta $%
{\footnotesize \ and }$\gamma ${\footnotesize , }$J_{2}${\footnotesize \
fixed to its minimum, average and maximum value respectively.} \\ 
{\footnotesize \qquad Additional constraints on }$\beta ${\footnotesize \
and }$\gamma ${\footnotesize \ (shaded region) can be taken from the
Nordtvedt effect and the L.L.R. data} \\ 
{\footnotesize (see (\ref{estimated value of beta from LLR and Nordtvedt
effect}) and (\ref{estimated value of gamma from L.L.R. data})).} \\ 
{\footnotesize They allow to determine a portion of the ellipses which
constitute the allowed parameter space for }$\beta ${\footnotesize \ and }$%
\gamma ${\footnotesize .}
\end{tabular}
\end{tabular}

\end{landscape}

\pagebreak

\begin{landscape}

\[
\begin{tabular}{c}
\hline\hline
\begin{tabular}{c}
- Table 2 - \\ 
Estimated values of the Solar quadrupole moment $J_{2}$ from solar
observations and solar modeling
\end{tabular}
\\ 
\begin{tabular}[t]{l|l|l|l|l}
\hline\hline
Year & References & Method & $J_{2}$ & Critics \\ \hline\hline
&  &  &  &  \\ 
$
\begin{tabular}[t]{l}
{\footnotesize 1890-1902}
\end{tabular}
$ & 
\begin{tabular}[t]{l}
{\footnotesize \cite{Ambronn 1905 J2 measurement with an heliometer}}$^{*}$
\\ 
{\footnotesize \cite{Wittmann 1987 Solar diameter and its variability}}$^{*}$%
\end{tabular}
& $
\begin{array}[t]{l}
\text{{\footnotesize Direct ground based observation of the solar oblateness}%
} \\ 
\text{{\footnotesize at G\"{o}ttingen (heliometer).}}
\end{array}
$ & 
\begin{tabular}[t]{l}
$\leq 4.4\ 10^{-6}$%
\end{tabular}
& 
\begin{tabular}[t]{l}
{\footnotesize a, c, d}
\end{tabular}
\\ \hline
$
\begin{tabular}[t]{l}
{\footnotesize 1891}
\end{tabular}
$ & 
\begin{tabular}[t]{l}
{\footnotesize \cite{Wittmann 1987 Solar diameter and its variability}}$^{*}$%
\end{tabular}
& 
\begin{tabular}[t]{l}
{\footnotesize Rotational theory of the Sun by Tisserand}
\end{tabular}
& 
\begin{tabular}[t]{l}
$<14\ 10^{-6}$%
\end{tabular}
&  \\ \hline
$
\begin{tabular}[t]{l}
{\footnotesize 1909}
\end{tabular}
$ & 
\begin{tabular}[t]{l}
{\footnotesize \cite{Wittmann 1987 Solar diameter and its variability}}$^{*}$%
\end{tabular}
& 
\begin{tabular}[t]{l}
{\footnotesize Rotational theory of the Sun by Moulton}
\end{tabular}
& 
\begin{tabular}[t]{l}
$<20\ 10^{-6}$%
\end{tabular}
&  \\ \hline
\begin{tabular}[t]{l}
{\footnotesize 1966}
\end{tabular}
& 
\begin{tabular}[t]{l}
{\footnotesize \cite{Dicke 1967 estimation J2 trops grande}}$^{*}$ \\ 
{\footnotesize \cite{Dicke 1974 The oblateness of the Sun}} \\ 
{\footnotesize \cite{Dicke 1986 Variable oblateness of the Sun}}$^{*}$%
\end{tabular}
& $
\begin{array}[t]{l}
\text{{\footnotesize Direct ground based observation of the solar oblateness}%
} \\ 
\text{{\footnotesize at Princeton}} \\ 
\text{{\footnotesize (integrated flux from inside till outside the limb).}}
\end{array}
$ & 
\begin{tabular}[t]{l}
$(2.37\pm 0.23)\ 10^{-5}$ \\ 
$(2.47\pm 0.23)\ 10^{-5}$%
\end{tabular}
& 
\begin{tabular}[t]{l}
{\footnotesize a, d, e} \\ 
{\footnotesize f, g, i}
\end{tabular}
\\ \hline
& 
\begin{tabular}{l}
{\footnotesize \cite{Goldreich 1968 Theoretical upper bound to the solar
oblateness}}
\end{tabular}
& $
\begin{array}[t]{l}
\text{{\footnotesize Theory of the solar figure obtained from a rotational}}
\\ 
\text{{\footnotesize law (based upon stability criteria under differential}}
\\ 
\text{{\footnotesize rotation and contemporary surface rotation}} \\ 
\text{{\footnotesize observations) plus a contemporary density model that}}
\\ 
\text{{\footnotesize are integrated from the center of the Sun }{\small till
its}} \\ 
\text{{\small surface, in order to derive }}\varepsilon \text{{\small \ at
the }{\footnotesize surface. It is}} \\ 
\text{{\footnotesize further constrained by a solar evolution model.}}
\end{array}
$ & 
\begin{tabular}{l}
$\leq 7.96\ 10^{-5}$%
\end{tabular}
& 
\begin{tabular}[t]{l}
{\footnotesize k, l, n,} \\ 
{\footnotesize v}
\end{tabular}
\\ \hline
\begin{tabular}[t]{l}
{\footnotesize 1972} \\ 
{\footnotesize 1973}
\end{tabular}
& 
\begin{tabular}[t]{l}
{\footnotesize \cite{Hill 1974 Solar oblateness excess brightness}}$^{*}$ \\ 
{\footnotesize \cite{Hill 1975 Intrinsic visual oblateness}}
\end{tabular}
& $
\begin{array}[t]{l}
\text{{\footnotesize Direct ground based observation of the solar oblateness}%
} \\ 
\text{{\footnotesize using a F.F.T. edge definition, during periods}} \\ 
\text{{\footnotesize of reduced excess brightness (diameter measurement}} \\ 
\text{{\footnotesize and excess equatorial brightness monitoring).}}
\end{array}
$ & 
\begin{tabular}{l}
$(9.72\pm 43.4)\ 10^{-7}$%
\end{tabular}
& 
\begin{tabular}[t]{l}
{\footnotesize a, d, f,} \\ 
{\footnotesize g}
\end{tabular}
\\ \hline
{\footnotesize (...)} &  &  &  &  \\ \hline
\end{tabular}
\end{tabular}
\]

\end{landscape}

\pagebreak

\begin{landscape}

\[
\begin{tabular}[t]{l|l|l|l|l}
\hline\hline
Year & References & Method & $J_{2}$ & Critics \\ \hline\hline
{\footnotesize (...)} &  &  &  &  \\ \hline
\begin{tabular}{l}
{\footnotesize 1979}
\end{tabular}
& 
\begin{tabular}{l}
{\footnotesize \cite{Gough 1982 Internal rotation and J2}}
\end{tabular}
& $
\begin{array}[t]{l}
\text{{\footnotesize Ground based observation of modes frequency}} \\ 
\text{{\footnotesize splittings in solar oscillation data allow to infer}}
\\ 
\text{{\footnotesize an internal radial }{\small rotation law from which to
deduce }}J_{2}\text{{\small .}}
\end{array}
$ & $
\begin{tabular}[t]{l}
$\geq 1.2\ 10^{-6}$ \\ 
{\footnotesize or} \\ 
$\sim 3.6\ 10^{-6}$%
\end{tabular}
$ & 
\begin{tabular}[t]{l}
{\footnotesize a, k, l,} \\ 
{\footnotesize p, t, u} \\ 
{\footnotesize v, x}
\end{tabular}
\\ \hline
\begin{tabular}{l}
{\footnotesize 1979}
\end{tabular}
& 
\begin{tabular}{l}
{\footnotesize \cite{Hill 1982 J2 from rotational splitting}}
\end{tabular}
& $
\begin{array}[t]{l}
\text{{\footnotesize Ground based observation of modes frequency}} \\ 
\text{{\footnotesize splittings in solar oscillation data allow to infer}}
\\ 
\text{{\footnotesize an internal radial }{\small rotation law from which to
deduce }}J_{2}\text{{\small .}}
\end{array}
$ & 
\begin{tabular}{l}
$(5.5\pm 1.3)\ 10^{-6}$%
\end{tabular}
& 
\begin{tabular}[t]{l}
{\footnotesize a, k, l,} \\ 
{\footnotesize p, t, u} \\ 
{\footnotesize v, w}
\end{tabular}
\\ \hline
& 
\begin{tabular}{l}
{\footnotesize \cite{Ulrich 1981 Solar gravitational figure}}
\end{tabular}
& $
\begin{array}[t]{l}
\text{{\footnotesize Theory of the solar figure obtained from a rotational
law}} \\ 
\text{{\footnotesize (based upon a differential rotation model and surface}}
\\ 
\text{{\footnotesize rotation observations) plus a density model that are}}
\\ 
\text{{\footnotesize integrated from the center of the Sun till its }{\small %
surface, in}} \\ 
\text{{\small order to derive }}J_{2}\text{{\small \ at the}{\footnotesize %
surface.}}
\end{array}
$ & 
\begin{tabular}{l}
$(1.25\pm 0.25)\ 10^{-7}$%
\end{tabular}
& 
\begin{tabular}[t]{l}
{\footnotesize k, l, n}
\end{tabular}
\\ \hline
& 
\begin{tabular}{l}
{\footnotesize \cite{Kislik 1983 On the solar oblateness}}
\end{tabular}
& $
\begin{array}[t]{l}
\text{{\footnotesize Theory of the solar figure obtained from a rotational
law}} \\ 
\text{{\footnotesize based upon rigid body like rotation, surface rotation}}
\\ 
\text{{\footnotesize observations and a homogenuous density model.}} \\ 
\text{{\footnotesize This provides an upper limit for }}J_{2}\text{{\small \
at the surface.}}
\end{array}
$ & 
\begin{tabular}{l}
$<1.08\ 10^{-5}$%
\end{tabular}
&  \\ \hline
\begin{tabular}{l}
{\footnotesize 1979}
\end{tabular}
& 
\begin{tabular}{l}
{\footnotesize \cite{Campbell 1983 Quadrupole moment and perihelion
precession}}
\end{tabular}
& $
\begin{array}[t]{l}
\text{{\footnotesize Ground based observation of modes frequency splittings}}
\\ 
\text{{\footnotesize in solar oscillation data allow to infer an internal
radial}} \\ 
\text{{\small rotation law from which to deduce }}J_{2}\text{{\small .}}
\end{array}
$ & $
\begin{tabular}[t]{l}
$\geq 1.6\ 10^{-6}$ \\ 
{\footnotesize or} \\ 
$\sim 5.0\ 10^{-6}$%
\end{tabular}
$ & 
\begin{tabular}[t]{l}
{\footnotesize a, p, t,} \\ 
{\footnotesize u, v, x}
\end{tabular}
\\ \hline
\begin{tabular}{l}
{\footnotesize 1983}
\end{tabular}
& 
\begin{tabular}[t]{l}
{\footnotesize \cite{Dicke 1985 Oblateness of the Sun in 1983}} \\ 
{\footnotesize \cite{Dicke 1986 Variable oblateness of the Sun}}$^{*}$ \\ 
{\footnotesize \cite{Dicke 1987 Solar oblateness variable?}}$^{*}$%
\end{tabular}
& $
\begin{array}[t]{l}
\text{{\footnotesize Direct ground based observation of the solar oblateness}%
} \\ 
\text{{\footnotesize during periods of reduced excess brightness, at M}}^{%
\text{{\footnotesize t}}}\text{{\footnotesize \ Wilson.}} \\ 
\text{{\footnotesize (integrated flux from inside till outside the limb).}}
\end{array}
$ & 
\begin{tabular}{l}
$(7.92\pm 0.972)\ 10^{-6}$%
\end{tabular}
& 
\begin{tabular}[t]{l}
{\footnotesize a, d, f,} \\ 
{\footnotesize g, i, j}
\end{tabular}
\\ \hline
{\footnotesize (...)} &  &  &  &  \\ \hline
\end{tabular}
\]

\end{landscape}

\pagebreak

\begin{landscape}

\[
\begin{tabular}[t]{l|l|l|l|l}
\hline\hline
Year & References & Method & $J_{2}$ & Critics \\ \hline\hline
{\footnotesize (...)} &  &  &  &  \\ \hline
\begin{tabular}[t]{l}
{\footnotesize 1984}
\end{tabular}
& 
\begin{tabular}{l}
{\footnotesize \cite{Brown 1989 J2 and p mode frequency splitting}}
\end{tabular}
& $
\begin{array}[t]{l}
\text{{\footnotesize Ground based observation of p-modes frequency}} \\ 
\text{{\footnotesize splittings in solar oscillation data allow to infer an}}
\\ 
\text{{\footnotesize internal angular }{\small rotation law from which to
deduce }}J_{2}.
\end{array}
$ & 
\begin{tabular}{l}
$(1.7\pm 10\%)\ 10^{-7}$%
\end{tabular}
& 
\begin{tabular}[t]{l}
{\footnotesize a, o, p,} \\ 
{\footnotesize t, u}
\end{tabular}
\\ \hline
\begin{tabular}[t]{l}
{\footnotesize 1984}
\end{tabular}
& 
\begin{tabular}{l}
{\footnotesize \cite{Duvall 1984 Internal Rotation of the Sun}}
\end{tabular}
& $
\begin{array}[t]{l}
\text{{\footnotesize Ground based observation of p-modes frequency splittings%
}} \\ 
\text{{\footnotesize in solar oscillation data (...)}}
\end{array}
$ & 
\begin{tabular}{l}
$(1.7\pm 0.4)\ 10^{-7}$%
\end{tabular}
& 
\begin{tabular}[t]{l}
{\footnotesize a, u, v}
\end{tabular}
\\ \hline
\begin{tabular}[t]{l}
{\footnotesize 1984}
\end{tabular}
& 
\begin{tabular}[t]{l}
{\footnotesize \cite{Dicke 1986 Variable oblateness of the Sun}}$^{*}$ \\ 
{\footnotesize \cite{Dicke 1987 Solar oblateness variable?}}$^{*}$%
\end{tabular}
& $
\begin{array}[t]{l}
\text{{\footnotesize Direct ground based observation of the solar oblateness}%
} \\ 
\text{{\footnotesize during periods of reduced excess brightness}{\small .}}
\\ 
\text{{\footnotesize (integrated flux from inside till outside the limb).}}
\end{array}
$ & 
\begin{tabular}{l}
$(-1.53\pm 2.36)\ 10^{-6}$%
\end{tabular}
& 
\begin{tabular}[t]{l}
{\footnotesize a, d, f,} \\ 
{\footnotesize g, j}
\end{tabular}
\\ \hline
\begin{tabular}[t]{l}
{\footnotesize 1985}
\end{tabular}
& 
\begin{tabular}{l}
{\footnotesize \cite{Dicke 1987 Solar oblateness variable?}}$^{*}$%
\end{tabular}
& $
\begin{array}[t]{l}
\text{{\footnotesize Direct ground based observation of the solar oblateness}%
} \\ 
\text{{\footnotesize during periods of reduced excess brightness.}} \\ 
\text{{\footnotesize (integrated flux from inside till outside the limb).}}
\end{array}
$ & 
\begin{tabular}{l}
$(4.72\pm 1.53)\ 10^{-6}$%
\end{tabular}
& 
\begin{tabular}[t]{l}
{\footnotesize a, d, f,} \\ 
{\footnotesize g, j}
\end{tabular}
\\ \hline
\begin{tabular}[t]{l}
{\footnotesize 1986}
\end{tabular}
& 
\begin{tabular}{l}
{\footnotesize \cite{Bursa 1986 Sun's flattening and influence on planetary
orbits}}
\end{tabular}
& 
\begin{tabular}[t]{l}
{\footnotesize Limits on the solar oblateness from the theory of solar} \\ 
{\footnotesize figure given by Roche's and MacLaurin's models.} \\ 
{\footnotesize The upper limit of a heavy core is taken to infer }$J_{2}$%
{\footnotesize .}
\end{tabular}
& 
\begin{tabular}{l}
$\leq 1.1\ 10^{-5}$%
\end{tabular}
&  \\ \hline
\begin{tabular}[t]{l}
{\footnotesize 1989}
\end{tabular}
& 
\begin{tabular}[t]{l}
{\footnotesize \cite{Delache 1994 Valeur de J2}}
\end{tabular}
& $
\begin{array}[t]{l}
\text{{\footnotesize Analysis of T. Brown's new helioseismic data}}
\end{array}
$ & 
\begin{tabular}[t]{l}
$(7.7\pm 2.1)\ 10^{-6}$%
\end{tabular}
& 
\begin{tabular}{l}
{\footnotesize a, p, t,} \\ 
{\footnotesize u}
\end{tabular}
\\ \hline
\begin{tabular}[t]{l}
{\footnotesize 1990}
\end{tabular}
& 
\begin{tabular}[t]{l}
{\footnotesize \cite{Maier 1992 preliminary results of SDS}}$^{*}$%
\end{tabular}
& $
\begin{array}[t]{l}
\text{{\footnotesize Solar Disk Sextant (S.D.S.): a baloon born experiment}}
\\ 
\text{{\footnotesize indirectly measuring the solar angular diameter at}} \\ 
\text{{\footnotesize a variety of orientations using the F.F.T. edge
definition}} \\ 
J_{2}\text{{\small \ is then evaluated from the }{\footnotesize infered
solar oblateness}} \\ 
\text{{\footnotesize and from solar surface angular rotation data.}}
\end{array}
$ & 
\begin{tabular}[t]{l}
$(+1.68\pm 5.70)\ 10^{-6}$%
\end{tabular}
& 
\begin{tabular}[t]{l}
{\footnotesize b, d, f,} \\ 
{\footnotesize g, y, z,} \\ 
{\footnotesize aa, bb,} \\ 
{\footnotesize cc, dd,} \\ 
{\footnotesize ee}
\end{tabular}
\\ \hline
{\footnotesize (...)} &  &  &  &  \\ \hline
\end{tabular}
\]

\end{landscape}

\pagebreak

\begin{landscape}

\[
\begin{tabular}[t]{l|l|l|l|l}
\hline\hline
Year & References & Method & $J_{2}$ & Critics \\ \hline\hline
{\footnotesize (...)} &  &  &  &  \\ \hline
\begin{tabular}[t]{l}
{\footnotesize 1992}
\end{tabular}
& 
\begin{tabular}[t]{l}
{\footnotesize \cite{Sofia 1994 SDS experiment}}
\end{tabular}
& $
\begin{array}[t]{l}
\text{{\footnotesize Solar Disk Sextant (S.D.S.) (...)}}
\end{array}
$ & 
\begin{tabular}{l}
$(0.3\pm 0.6)\ 10^{-6}$%
\end{tabular}
& 
\begin{tabular}[t]{l}
{\footnotesize b, d, f,} \\ 
{\footnotesize g, y, z,} \\ 
{\footnotesize aa, dd}
\end{tabular}
\\ \hline
\begin{tabular}[t]{l}
{\footnotesize 1992-1994}
\end{tabular}
& 
\begin{tabular}[t]{l}
{\footnotesize \cite{Elsworth 1995 Slow rotation ot the Sun's interior}}
\end{tabular}
& 
\begin{tabular}[t]{l}
{\footnotesize Ground based observation of p-mode frequency splittings} \\ 
{\footnotesize in solar oscillation data obtained from the Birmingham} \\ 
{\footnotesize Solar Oscillation Network (BiSON) allow to infermm an} \\ 
{\footnotesize internal rotaion law from which to deduce }$J_{2}$%
{\footnotesize .}
\end{tabular}
& 
\begin{tabular}[t]{l}
$(2.0\pm 0.5)\ 10^{-7}$%
\end{tabular}
& 
\begin{tabular}[t]{l}
{\footnotesize a, l, p,} \\ 
{\footnotesize t, ii,} \\ 
{\footnotesize jj}
\end{tabular}
\\ \hline
\begin{tabular}[t]{l}
{\footnotesize 1992} \\ 
{\footnotesize 1994}
\end{tabular}
& 
\begin{tabular}{l}
{\footnotesize \cite{Lydon 1996 Solar quadrupole moment}}
\end{tabular}
& $
\begin{array}[t]{l}
\text{{\footnotesize Solar Disk Sextant (S.D.S.) (...)}}
\end{array}
$ & 
\begin{tabular}{l}
$(1.8\pm 5.1)\ 10^{-7}$%
\end{tabular}
& 
\begin{tabular}[t]{l}
{\footnotesize b, d, f,} \\ 
{\footnotesize g, y, z,} \\ 
{\footnotesize aa, dd} \\ 
{\footnotesize ff, gg}
\end{tabular}
\\ \hline
\begin{tabular}[t]{l}
{\footnotesize 1990} \\ 
{\footnotesize 1991} \\ 
{\footnotesize 1992} \\ 
{\footnotesize 1994}
\end{tabular}
& 
\begin{tabular}{l}
{\footnotesize \cite{Paterno1996 Sun rotation}}
\end{tabular}
& $
\begin{array}[t]{l}
\text{{\footnotesize Solar Disk Sextant (S.D.S.) (1992 and 1994) lead to a}}
\\ 
\text{{\footnotesize measurement of the oblateness (...) which is used with a%
}} \\ 
\text{{\small rotation model in order to evaluate }}J_{2}\text{{\small . }%
{\footnotesize The surface}} \\ 
\text{{\footnotesize rotation model is constrained by ground based}} \\ 
\text{{\footnotesize observations of acoustic p-modes frequency}} \\ 
\text{{\footnotesize splittings from either the helioseismic network IRIS
(1991}} \\ 
\text{{\footnotesize -1992) or BiSON (1992-1994).}}
\end{array}
$ & $
\begin{array}[t]{l}
(2.22\pm 0.1)\ 10^{-7} \\ 
\text{{\footnotesize IRIS}} \\ 
\updownarrow \\ 
(2.08\pm 0.14)\ 10^{-7} \\ 
\text{{\footnotesize BISON}}
\end{array}
$ & 
\begin{tabular}[t]{l}
{\footnotesize S.D.S:} \\ 
{\footnotesize b, d, f,} \\ 
{\footnotesize g, y, z,} \\ 
{\footnotesize aa, dd, ff;} \\ 
{\footnotesize BiSON/} \\ 
{\footnotesize IRIS:} \\ 
{\footnotesize a, p, hh,} \\ 
{\footnotesize ii}
\end{tabular}
\\ \hline
{\footnotesize (...)} &  &  &  &  \\ \hline
\end{tabular}
\]

\end{landscape}

\pagebreak

\begin{landscape}

\[
\begin{tabular}[t]{l|l|l|l|l}
\hline\hline
Year & References & Method & $J_{2}$ & Critics \\ \hline\hline
{\footnotesize (...)} &  &  &  &  \\ \hline
\begin{tabular}[t]{l}
{\footnotesize 1993} \\ 
{\footnotesize 1994}
\end{tabular}
& 
\begin{tabular}[t]{l}
{\footnotesize \cite{Rosch 1996 A new estimate of J2}}
\end{tabular}
& $
\begin{array}[t]{l}
\text{{\footnotesize Direct ground based observation of the solar oblateness}%
} \\ 
\text{{\footnotesize during periods of reduced excess brightness using the}}
\\ 
\text{{\footnotesize distance between both inflexion points of the limb}} \\ 
\text{{\footnotesize profile (scanning heliometer provides diameter}} \\ 
\text{{\footnotesize measurements and excess brightness monitoring).}}
\end{array}
$ & 
\begin{tabular}[t]{l}
$(2.57\pm 2.36)\ 10^{-6}$%
\end{tabular}
& 
\begin{tabular}[t]{l}
{\footnotesize a, d, f,} \\ 
{\footnotesize g, h}
\end{tabular}
\\ \hline
\begin{tabular}[t]{l}
{\footnotesize 1995} \\ 
{\footnotesize 1996}
\end{tabular}
& 
\begin{tabular}[t]{l}
{\footnotesize \cite{Pijpers 1998 quadrupole moment}}
\end{tabular}
& $
\begin{array}[t]{l}
\text{{\footnotesize Ground based observation of frequency splittings in
solar}} \\ 
\text{{\footnotesize oscillation data obtained from the Global Oscillation
Network}} \\ 
\text{{\footnotesize Group (GONG)}{\small \ }{\footnotesize (1995-1996) or
space observations of}} \\ 
\text{{\footnotesize oscillations (``a'' coefficients) by the Solar
Heliospheric}} \\ 
\text{{\footnotesize Observatory (SoHO) (1996) allow to infer an internal
angular}} \\ 
\text{{\footnotesize rotation law from which }{\small to deduce }}J_{2}.
\end{array}
$ & $
\begin{array}[t]{l}
(2.14\pm 0.09)\ 10^{-7} \\ 
\text{{\footnotesize GONG}} \\ 
\updownarrow \\ 
(2.23\pm 0.09)\ 10^{-7} \\ 
\text{{\footnotesize SoHO}} \\ 
\Rightarrow \\ 
(2.18\pm 0.06)\ 10^{-7} \\ 
\text{{\footnotesize mean value}}
\end{array}
$ & 
\begin{tabular}[t]{l}
{\footnotesize a (GONG),} \\ 
{\footnotesize k, p}
\end{tabular}
\\ \hline
\begin{tabular}[t]{l}
{\footnotesize 1996}
\end{tabular}
& 
\begin{tabular}[t]{l}
{\footnotesize \cite{Rozelot 1997 Upper bound for J2}}
\end{tabular}
& $
\begin{array}[t]{l}
\text{{\footnotesize Direct ground based observation of the solar oblateness}%
} \\ 
\text{{\footnotesize during periods of reduced excess brightness using the}}
\\ 
\text{{\footnotesize distance between both inflexion points of the limb}} \\ 
\text{{\footnotesize profile (scanning heliometer provides diameter}} \\ 
\text{{\footnotesize measurements and excess brightness monitoring).}}
\end{array}
$ & 
\begin{tabular}{l}
$(7.57\pm 15)\ 10^{-7}$%
\end{tabular}
& 
\begin{tabular}[t]{l}
{\footnotesize a, d, f,} \\ 
{\footnotesize g, h}
\end{tabular}
\\ \hline
{\footnotesize (...)} &  &  &  &  \\ \hline
\end{tabular}
\]

\end{landscape}

\pagebreak

\begin{landscape}

\[
\begin{tabular}[t]{l|l|l|l|l}
\hline\hline
Year & References & Method & $J_{2}$ & Critics \\ \hline\hline
{\footnotesize (...)} &  &  &  &  \\ \hline
\begin{tabular}[t]{l}
{\footnotesize 1996} \\ 
{\footnotesize 1997}
\end{tabular}
& 
\begin{tabular}[t]{l}
{\footnotesize \cite{Kuhn 1998 Comparison between SoHO and ground-based
results on the oblateness}}$^{*}$%
\end{tabular}
& 
\begin{tabular}[t]{l}
{\footnotesize Space observation, by SoHO satellite, of the Sun's full limb}
\\ 
{\footnotesize position (Michelson Doppler Imager -M.D.I.- experiment)} \\ 
{\footnotesize and brightness allow to infer an oblateness from which to} \\ 
{\footnotesize deduce }$J_{2}$ {\footnotesize by legendre polynomial fit to
the observed limb.} \\ 
{\footnotesize (During periods of reduced solar magnetic activity).}
\end{tabular}
& $
\begin{array}[t]{l}
(-12.5\pm 20.1)\ 10^{-7} \\ 
\updownarrow  \\ 
(-18.2\pm 17.6)\ 10^{-7} \\ 
\Rightarrow  \\ 
(-16.8\pm 17.3)\ 10^{-7} \\ 
\text{{\footnotesize mean value of 1996-1997}}
\end{array}
$ & 
\begin{tabular}[t]{l}
{\footnotesize d, f, q,} \\ 
{\footnotesize r}
\end{tabular}
\\ \hline
& 
\begin{tabular}[t]{l}
{\footnotesize \cite{Rozelot 1998 new results on solar oblateness}}
\end{tabular}
& 
\begin{tabular}[t]{l}
{\footnotesize Constraints on }$J_{2}${\footnotesize \ from the accurate
knowledge of the} \\ 
{\footnotesize moon's physical librations, for which the L.L.R. data} \\ 
{\footnotesize reach accuracies at the milli- arcsec level.}
\end{tabular}
& 
\begin{tabular}{l}
$\leq 3\ 10^{-6}$%
\end{tabular}
& 
\begin{tabular}[t]{l}
{\footnotesize s}
\end{tabular}
\\ \hline
& 
\begin{tabular}[t]{l}
{\footnotesize \cite{Roxburg 2000 Gravitational multipole moment of the Sun}}
\end{tabular}
& $
\begin{array}[t]{l}
\text{{\footnotesize Theory of the solar figure obtained from a rotational
law}} \\ 
\text{{\footnotesize (based upon a differential rotation model -deduced from}%
} \\ 
\text{{\footnotesize helioseismic inversion- and surface rotation
observations)}} \\ 
\text{{\footnotesize plus a density model that are integrated from the center%
}} \\ 
\text{{\footnotesize of the Sun till its surface, to }{\small derive }}J_{2}%
\text{{\small \ at the surface.}}
\end{array}
$ & 
\begin{tabular}{l}
$(2.2125\pm 0.0075)\ 10^{-7}$%
\end{tabular}
& 
\begin{tabular}[t]{l}
{\footnotesize k, l}
\end{tabular}
\\ \hline
& 
\begin{tabular}[t]{l}
{\footnotesize \cite{Godier-Rozelot 1999a rotational potential}} \\ 
{\footnotesize \cite{Godier-Rozelot 1999b Relationship J2 and layers of the
Sun}} \\ 
{\footnotesize \cite{Godier-Rozelot 2000 Solar Oblateness}}
\end{tabular}
& $
\begin{array}[t]{l}
\text{{\footnotesize Theory of the solar figure obtained from a rotational
law}} \\ 
\text{{\footnotesize (based upon a differential rotation model -deduced from}%
} \\ 
\text{{\footnotesize helioseismic data and p-modes frequency splittings}} \\ 
\text{{\footnotesize obtained by SoHO- and surface rotation observations)}}
\\ 
\text{{\footnotesize plus a density model that are integrated from the center%
}} \\ 
\text{{\footnotesize of the Sun till its }{\small surface, in order to
derive }}J_{2}\text{{\small \ at }{\footnotesize the}} \\ 
\text{{\footnotesize surface.}}
\end{array}
$ & 
\begin{tabular}{l}
$(1.6\pm 0.4)\ 10^{-7}$%
\end{tabular}
& 
\begin{tabular}[t]{l}
{\footnotesize k, l, m}
\end{tabular}
\\ \hline
& 
\begin{tabular}[t]{l}
{\footnotesize \cite{Kuhn 2001 Private communication }}
\end{tabular}
& 
\begin{tabular}[t]{l}
{\footnotesize Reanalysis of observations.}
\end{tabular}
& 
\begin{tabular}{l}
$2.22\ 10^{-7}$%
\end{tabular}
&  \\ \hline\hline
\end{tabular}
\]

\end{landscape}

\pagebreak

Table 2. {\small To each reference corresponds an estimated value of the
solar quadru-}\newline
{\small pole moment, the method used to obtain this estimation (solar
observations, solar modeling), and some critics we formulate in regards to
the method. The year given in the table is the date the observations were
made (not the date of the publication).}\newline
{\small Notice that some authors}\footnote{%
For those authors mentioned with an asterix in the table, the value of $%
J_{2} $ as been inferred from their given value of the oblateness, $%
\varepsilon $, or excess equatorial radius, $\Delta r$.}{\small \ only
provide the value of solar equatorial excess radius (}$\Delta r\equiv
r_{equ}-r_{pol}${\small ) in their article. We thus inferred the solar
quadrupole moment (}$J_{2}${\small ) using the following formula \cite
{Rozelot 1996 Measure of sun's changing sizes} }$J_{2}=2/3\ \left( \Delta
r-\delta r\right) /r_{0}${\small , where }$\delta r=7.8\pm 2.1${\small \ arc
ms \cite{Rozelot 1997 Upper bound for J2}\textbf{\ }is the contribution to }$%
J_{2}${\small \ due to the surface rotation alone, }$\varepsilon \equiv
\Delta r/r_{0}${\small \ is the solar oblateness and }$r_{0}=9.6\ 10^{5}\ $%
{\small arc ms, the solar radius (i.e. the best sphere passing through }$%
r_{equ}${\small \ and }$r_{pol}$ {\small \cite{Rozelot 2001 solar radius
definition}).}\newline
{\small The critics or remarks made are the following ones:\newline
(a) Ground based experiments are subject to all kinds of atmospheric
perturbations that have to be modeled.}\newline
{\small (b) Balloon flights are still subject to some differential
refraction due to residual atmosphere (instability) and problems linked to
the stability of the pointing instruments.}\newline
{\small (c) The maximum value for }$\Delta r${\small \ is taken, as the
measured minimum leads to the erroneous prediction of an oblong Sun.}\newline
{\small (d) Observations of the oblateness have to be done only during
periods of reduced excess brightness in order to be able to deduce the
intrinsic visual oblateness from the apparent oblateness obtained with
whichever edge definition... until the mechanisms of excess brightness are
understood and proper models exists for it.}\newline
{\small (e) Did not take into account the solar surface }$\nabla T{}^{\circ
} ${\small \ which could lead to a difference in brightness
indistinguishable from a geometrical oblateness.}\newline
{\small (f) The choice of the edge of the sun's definition profoundly
influences the sensitivity to excess brightness.}\newline
{\small (The F.F.T. -Finite Fourier Transform- edge definition is highly
sensitive to the limb darkening shape, but this allow a simultaneous
sensitive monitoring of the excess brightness, and detecting local/global
active regions without reliance on solar atmosphere models or other
observations.)}\newline
{\small This leads to discrepancies among the different results obtained for
oblateness measurements made during the same period (even with the same
instrument!) but using different edge definitions.}\newline
{\small (g) The choice of the edge of the sun's definition profoundly
influences the apparent displacement of the Sun's edge attributable to
atmospheric seeing.}\newline
{\small (The F.F.T. edge definition is less sensitive to this effect than
the Dicke Goldenberg integral edge definition.)}\newline
{\small (h) Difficulty to correct for the shift of the inflection point.}%
\newline
{\small (i) The stated error on }$\Delta r${\small \ for the 1966 and 1983
experiment is a formal standard deviation. To make allowance for possible
seasonal variations in the locally induced atmospheric distortions, the
error bars should be increased, possibly to }$4${\small \ ms. The 1984 and
1985 results are already corrected for this error and thus the derived value
of }$J_{2}${\small . Notice that 1966 results have often been reinterpreted
by the authors leading to different conclusions (see \cite{Dicke 1976 Solar
distortion}).}\newline
{\small (j) Observations in 1983, 1984 and 1985 have been made with a
modified instrument (see \cite{Dicke 1985 Oblateness of the Sun in 1983}),
by comparison to the 1966 experiment of Dicke-Goldenberg, that automatically
excluded data that was contaminated by signals due to substantial facular
patches, as well as color dependent brightness signals. The possible
existence of a color independent brightness signal is however not taken into
account.}\newline
{\small (k) Dependent upon the solar density model.}\newline
{\small (l) Dependent upon the solar rotation model.}\newline
{\small (m) Assumes the same rotation rate for the core and for the
radiative zone; but allows }$\Omega (r,\theta )${\small \ to vary with the
latitude.}\newline
{\small (n) Uses a model of internal rotation which is assumed to be
uniformly differential (constant on cylinders) through out the convective
envelope, and non differential below the convective zone.}\newline
{\small (o) No reliable estimates for the uncertainties.}\newline
{\small (p) Helioseismic data are limited by the error bars to distances
above }$0.2\ R_{s}${\small , near the surface and near the poles..}\newline
{\small (q) A mean limb darkening function is used, while a more realistic
model should use a more complex function.}\newline
{\small (r) 1996 data set gives noisier results as it was obtained without
the active M.D.I. image stabilization system.}\newline
{\small (s) Simulations have been performed assuming }$J_{2}${\small \
constant.}\newline
{\small (t) It is difficult to correctly identify individual modes of
oscillation for all spatial scales. Moreover, the oscillations must be
adequately long lived.}\newline
{\small (u) Oscillations, as observed from the ground, can not provide a
good measure of the rotation rate near the solar pole, because
foreshortening limits the viewing region.}\newline
{\small (v) Assume that the rotational frequency is independent of the
latitude.}\newline
{\small (w) Maximum consistent with the stability of the Sun.}\newline
{\small (x) Some fits to the data produced rotation curves, }$\Omega (r)$%
{\small , that were highly unphysical. The given value for }$J_{2}${\small \
corresponds to the smoothest curve that fits the data.}\newline
{\small (y) A small quantity, the separation between 2 solar limb images, is
measured, instead of the full solar diameter. This enhances the precision
with respect to the techniques that measure the full diameter directly.}%
\newline
{\small (z) The instrument scale can be calibrated for any measurement.}%
\newline
{\small (aa) The quantity measured is located near the optical axis of the
instrument (unlike for direct diameter measurements), where the optical
system is optimal.}\newline
{\small (bb) No true perpendicular diameters are measured (polar and
equatorial). The resulting }$J_{2}${\small \ is thus probably less than its
real value.}\newline
{\small (cc) The resulting oblateness is }$\sim ${\small 30}${{}^{\circ }}$%
{\small \ offset from the polar-equator position.}\newline
{\small (dd) Solar photospheric }$T{}^{\circ }${\small \ and }$\nabla
T{}^{\circ }${\small \ may be a function of the activity solar cycle, and
so, the use of F.F.T. definitions into data reduction from the S.D.S
experiment would introduce systematic errors. }\newline
{\small (ee) Gravitational distortions (a non constant wedge angle) of the
instrument exist that were avoided in the next balloon flights (1992-1994).}%
\newline
{\small (ff) S.D.S. experiments were made on 2 days, 2 years apart
(1992-1994) rather than continuously over a period of many years. Moreover,
there were no observations of solar surface rotation available between
1992-1994. Thus, the large number of observations did not allow to lower the
uncertainties.}\newline
{\small (gg) Uses a simple model of internal rotation of the Sun as constant
angular rotation on cylinders or on cones.}\newline
{\small (hh) BiSON's helioseismic data imply a solar rotation law which is
not compatible with that inferred from IRIS's.}\newline
{\small (ii) The rotation model does not take into account helioseismologic
observations made at different latitudes.}\newline
{\small (jj) It was not possible to find a rotation model that reproduced
both the splitting data reported by \cite{Elsworth 1995 Slow rotation ot the
Sun's interior} and the data from the Big Bear Solar Observatory (B.B.S.O.)}%
\bigskip

Table 3. {\small This table follows table 1. To each reference corresponds
an inferred value of the solar quadrupole moment, in the setting of G.R.,
using the perihelion shift of Mercury.}\newline
{\small Notice that concerning references \cite{Krasinsky 1986 Relativistic
effects on planetary observations}, \cite{Krasinsky 1993 Motion of major
planets}, \cite{Pitjeva1993 Mercury toppography} and \cite{Pitjeva 2001
Modern Numerical ephemerides}, the value of the solar quadrupole moment is
calculated according to the formula given in the comments of table 1.}%
\pagebreak

\[
\begin{tabular}{c}
\hline
\begin{tabular}{c}
- Table 3 - \\ 
Inferred solar quadrupole moment \\ 
from the perihelion shift of Mercury, assuming G.R.
\end{tabular}
\\ \hline
\begin{tabular}[t]{l|l}
References & 
\begin{tabular}[t]{l}
Inferred Quadrupole Moment \\ 
$\left. 
\begin{array}{c}
J_{2} \\ 
(10^{-7})
\end{array}
\right. =\frac{\Delta \omega _{obs}-\Delta \omega _{0\ GR}}{\Delta \omega
_{0\ GR}\ 2.8218\emph{\ }10^{-4}}$%
\end{tabular}
\\ \hline
&  \\ 
\begin{tabular}{l}
{\small \cite{Newcomb 1895-1898 Perihelion advance and J2}}
\end{tabular}
& \multicolumn{1}{|c}{$\sim +${\small 32.1}} \\ 
\begin{tabular}{l}
{\small \cite{Clemence 1943 Perihelion advance}}
\end{tabular}
& \multicolumn{1}{|c}{$-${\small 11.6}$\pm ${\small 83.3}} \\ 
\begin{tabular}{l}
{\small \cite{Clemence 1947 Perihelion advance}}
\end{tabular}
& \multicolumn{1}{|c}{$-${\small 33.9}$\pm ${\small 79.2}} \\ 
\begin{tabular}{l}
{\small \cite{Duncombe 1958 Mercury (Planet)}}
\end{tabular}
& \multicolumn{1}{|c}{$+${\small 9.8}$\pm ${\small 36.3}} \\ 
\begin{tabular}{l}
{\small \cite{Wayman 1966 Determination of the Inertial frame of reference}}
\end{tabular}
& \multicolumn{1}{|c}{$+${\small 79.9}$\pm ${\small 33.8}} \\ 
\begin{tabular}{l}
{\small \cite{Shapiro 1972 Mercury perihelion advance and radar data}}
\end{tabular}
& \multicolumn{1}{|c}{$+${\small 13.9}$\pm ${\small 24.7}} \\ 
\begin{tabular}{l}
{\small \cite{Morrison 1975 Analysis of the transits of Mercury}}
\end{tabular}
& \multicolumn{1}{|c}{$-${\small 89.1}$\pm ${\small 41.2}} \\ 
\begin{tabular}{l}
{\small \cite{Shapiro 1976 equivalence principle}}
\end{tabular}
& \multicolumn{1}{|c}{$+${\small 10.6}$\pm ${\small 17.3}} \\ 
\begin{tabular}{l}
{\small \cite{Anderson 1978 Tests of GR using astrometric and radiometric
observations}}
\end{tabular}
& \multicolumn{1}{|c}{$+${\small 26.3}$\pm ${\small 16.5}} \\ 
$\left. 
\begin{array}{l}
\text{{\small \cite{Bretagnon 1982a Perihelion advance}}} \\ 
\text{{\small \cite{Narlikar 1985 N-body calculation}}}
\end{array}
\right\} $ & \multicolumn{1}{|c}{$+${\small 199.4}$\pm ${\small 4.1}} \\ 
$\left. 
\begin{array}{l}
\text{{\small \cite{Bretagnon 1982b Perihelion advance}}} \\ 
\text{{\small \cite{Rana 1987 motion of the node}}}
\end{array}
\right\} $ & \multicolumn{1}{|c}{$+${\small 187.1}$\pm ${\small 4.1}} \\ 
\begin{tabular}{l}
{\small \cite{Krasinsky 1986 Relativistic effects on planetary observations}:%
}
\end{tabular}
& \multicolumn{1}{|c}{} \\ 
$\quad 
\begin{array}[t]{l}
\text{{\small EPM1988}} \\ 
\text{{\small DE200}}
\end{array}
$ & \multicolumn{1}{|c}{$
\begin{array}[t]{c}
-\text{{\small 12.3}}\pm \text{{\small 10.0}} \\ 
-\text{{\small 17.1}}\pm \text{{\small 9.7}}
\end{array}
$} \\ 
\begin{tabular}{l}
{\small \cite{Rana 1987 motion of the node}}
\end{tabular}
& \multicolumn{1}{|c}{$+${\small 205.2}$\pm ${\small 7.4}} \\ 
\begin{tabular}{l}
{\small \cite{Anderson 1987 Ephemeris Mercury}}
\end{tabular}
& \multicolumn{1}{|c}{$-${\small 5.0}$\pm ${\small 16.5}} \\ 
\begin{tabular}{l}
{\small \cite{And 1991 IAU proc}}
\end{tabular}
& \multicolumn{1}{|c}{$-${\small 3.4}$\pm ${\small 16.5}} \\ 
\begin{tabular}{l}
{\small \cite{Anderson 1992 Singapore proc}}
\end{tabular}
& \multicolumn{1}{|c}{$+${\small 12.3}$\pm ${\small 11.5}} \\ 
\begin{tabular}{l}
{\small \cite{Krasinsky 1993 Motion of major planets}:}
\end{tabular}
& \multicolumn{1}{|c}{} \\ 
$\quad 
\begin{array}[t]{l}
\text{{\small EPM1988}} \\ 
\text{{\small DE200}}
\end{array}
$ & \multicolumn{1}{|c}{$
\begin{array}[t]{c}
+\text{{\small 0.33}}\pm \text{{\small 5.03}} \\ 
-\text{{\small 0.25}}\pm \text{{\small 5.03}}
\end{array}
$} \\ 
\begin{tabular}{l}
{\small \cite{Pitjeva1993 Mercury toppography}:}
\end{tabular}
& \multicolumn{1}{|c}{} \\ 
$\quad 
\begin{array}[t]{l}
\text{{\small EPM1988}} \\ 
\text{{\small DE200}}
\end{array}
$ & \multicolumn{1}{|c}{$
\begin{array}[t]{c}
-\text{{\small 1.40}}\pm \text{{\small 4.29}} \\ 
-\text{{\small 0.91}}\pm \text{{\small 4.29}}
\end{array}
$} \\ 
\begin{tabular}{l}
{\small \cite{Standish 2000 private communication}:}
\end{tabular}
& \multicolumn{1}{|c}{} \\ 
$\quad 
\begin{array}[t]{l}
\text{{\small DE405}}
\end{array}
$ & \multicolumn{1}{|c}{$
\begin{array}{l}
+\text{{\small 1.90}}\pm \text{{\small 0.16}}
\end{array}
$} \\ 
\begin{tabular}{l}
{\small \cite{Pitjeva 2001 Modern Numerical ephemerides}:}
\end{tabular}
& \multicolumn{1}{|c}{} \\ 
$\quad 
\begin{array}{l}
\text{{\small EPM2000}}
\end{array}
$ & \multicolumn{1}{|c}{$
\begin{array}{l}
+\text{{\small 2.453}}\pm \text{{\small 0.701}}
\end{array}
$}
\end{tabular}
\\ \hline
\end{tabular}
\]

\pagebreak

\[
\begin{tabular}{c}
\hline
\begin{tabular}{c}
- Table 4 - \\ 
Inferred solar quadrupole moment \\ 
from the perihelion shift of Icarus, assuming G.R.
\end{tabular}
\\ \hline
\begin{tabular}[t]{l|l}
References & 
\begin{tabular}[t]{l}
Inferred Quadrupole Moment \\ 
$J_{2}$ \\ 
(by fitting the parameters)
\end{tabular}
\\ \hline
& \multicolumn{1}{|c}{} \\ 
{\small \cite{Lieske 1969 Icarus and J2}} & \multicolumn{1}{|c}{
\begin{tabular}{l}
$\left( +1.8\pm 2.0\right) \ 10^{-5}$%
\end{tabular}
} \\ 
{\small \cite{Landgraf 1992 Estimation of J2 from Icarus}} & 
\multicolumn{1}{|c}{
\begin{tabular}[t]{l}
$\left( -0.65\pm 5.84\right) \ 10^{-6}$ \\ 
or $\leq 2\ 10^{-5}$%
\end{tabular}
}
\end{tabular}
\\ \hline
\end{tabular}
\]

Table 4. {\small To each reference corresponds an inferred value of the
solar quadrupole moment, in the setting of G.R., using the perihelion shift
of Icarus.}

\pagebreak

\begin{landscape}

\begin{tabular}[t]{l}
\FRAME{itbpF}{7.7418in}{4.0473in}{0in}{}{}{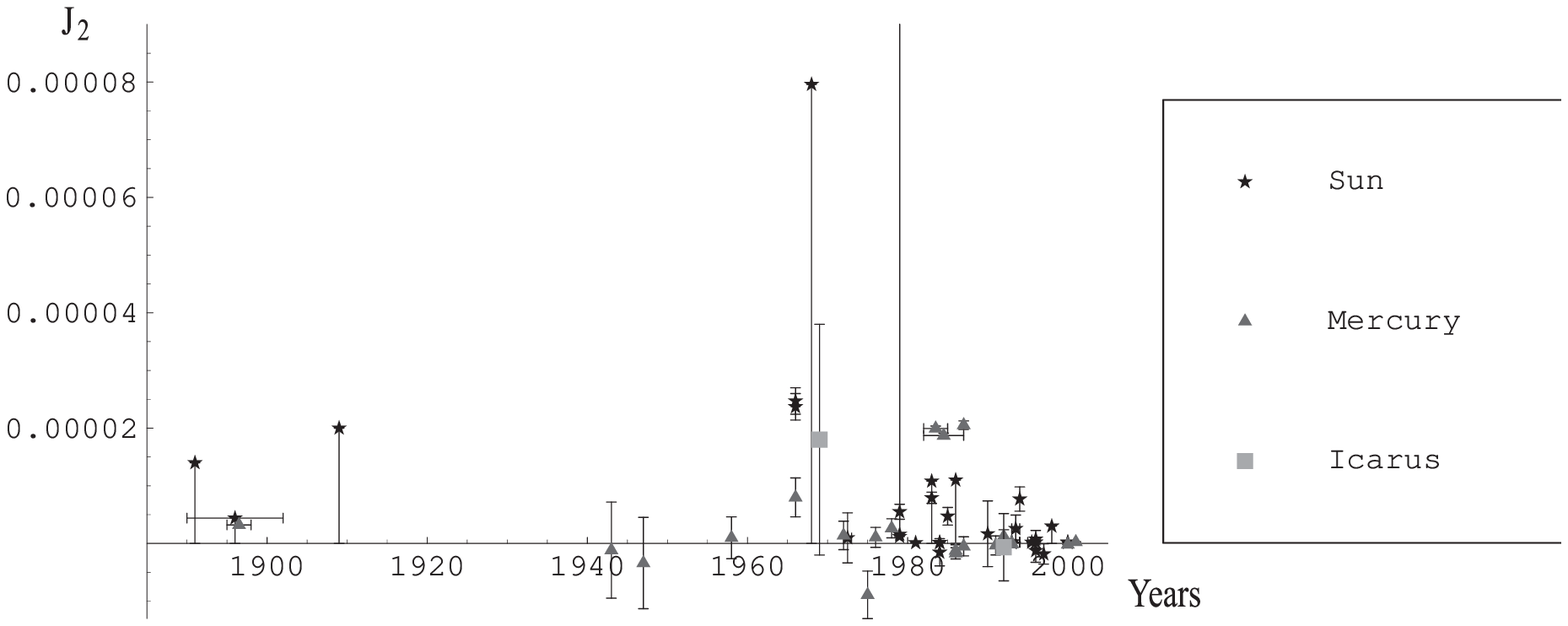}{%
\special{language "Scientific Word";type "GRAPHIC";display "ICON";valid_file
"F";width 7.7418in;height 4.0473in;depth 0in;original-width
0pt;original-height 0pt;cropleft "0.004714";croptop "1";cropright
"1.004714";cropbottom "0";filename
'figure4.eps';file-properties "XNPEU";}}
\\ 
Figure 2. {\small Different estimated values of the solar quadrupole moment, 
}$J_{2}${\small , versus the date when the respective observations} \\ 
{\small were made. There are 3 types data points: values estimated from
solar models and observations (table 2), values infered} \\ 
{\small from the perihelion shift of Mercury (table 3) and those obtained
from Icarus' (table 4).}
\end{tabular}

\end{landscape}

\pagebreak

\begin{landscape}

\[
\begin{tabular}{c}
\hline\hline
\begin{tabular}{c}
- Table 5 - \\ 
Proposed space missions dedicated to $\gamma $%
\end{tabular}
\\ \hline\hline
\begin{tabular}[t]{l|l|l|l}
References & Method & Mission & $
\begin{array}[t]{l}
\text{Expected} \\ 
\text{precision} \\ 
\text{on }\gamma 
\end{array}
$ \\ \hline
\begin{tabular}[t]{l}
{\small \cite{Bertotti 1998 Solar coronal plasma in doppler measurements}}
\\ 
{\small \cite{Iees 1999 Cassini Mission}}
\end{tabular}
& $
\begin{array}[t]{l}
\text{{\small doppler measurement of}} \\ 
\text{{\small the Solar gravitational deflection,}} \\ 
\text{{\small the first time this method is used}}
\end{array}
$ & 
\begin{tabular}[t]{l}
{\small Cassini launched in 1997,} \\ 
{\small experiment in 2002-2003}
\end{tabular}
& $10^{-4}-10^{-5}$ \\ \hline
\begin{tabular}[t]{l}
{\small \cite{Fi 1995 Gravity Probe B}, \cite{GAIA 2000 March study report}}
\\ 
{\small http://einstein.standford.edu}
\end{tabular}
& 
\begin{tabular}[t]{l}
{\small relativity gyroscope experiment,} \\ 
{\small geodetic precession measurement}
\end{tabular}
& 
\begin{tabular}[t]{l}
{\small Gravity Probe B (2002)}
\end{tabular}
& $6$\ $10^{-5}$ \\ \hline
$
\begin{array}[t]{l}
\text{{\small \cite{BepiColombo 2000 Study report}, \cite{Turyshev 1996
Mercury orbiter mission}}} \\ 
\begin{tabular}{l}
{\small http://www.estec.esa.nl/} \\ 
{\small spdwww/future/html/} \\ 
{\small meo2.htm}
\end{tabular}
\end{array}
$ & 
\begin{tabular}[t]{l}
{\small output of orbit determination,} \\ 
{\small time delay and doppler shift} \\ 
{\small measurements} \\ 
{\small (see section \ref{Mercury Orbiter})}
\end{tabular}
& $
\begin{array}[t]{l}
\text{{\small Mercury Orbiter, within}} \\ 
\text{{\small BepiColombo (2007/2009)}}
\end{array}
$ & $2.5\ 10^{-6}$ \\ \hline
$
\begin{array}[t]{l}
\text{{\small \cite{Reinhard 1999 SORT et autres}}} \\ 
\begin{tabular}{l}
{\small http://www.cnes.fr/WEB} \\ 
{\small \_UK/activities/index.htm,} \\ 
{\small see in ``Understanding the} \\ 
{\small universe'',``Fund. Phys.''}
\end{tabular}
\end{array}
$ & 
\begin{tabular}[t]{l}
{\small Projet d'Hologe Atomique par re-} \\ 
{\small froidissement d'Atomes en Orbite} \\ 
{\small (PHARAO clock), a swiss hydro-} \\ 
{\small gen maser clock to provide a long} \\ 
{\small term frequency standard, associated} \\ 
{\small with the IIS, to form the Atomic} \\ 
{\small Clock Ensemble in Space (ACES)}
\end{tabular}
& $
\begin{array}[t]{l}
\text{{\small International Space Station}} \\ 
\text{{\small (IIS) (2004/2005)}}
\end{array}
$ & $1\ 10^{-5}$ \\ \hline
$
\begin{tabular}{l}
$\text{{\small \cite{Reinhard 1999 SORT et autres}}}$%
\end{tabular}
$ & 
\begin{tabular}[t]{l}
{\small Time delay / light deflection measu-} \\ 
{\small rements (see section \ref{Mercury Orbiter})}
\end{tabular}
& $
\begin{array}[t]{l}
\text{{\small Solar Orbit Relativity Test}} \\ 
\text{{\small (SORT) (after 2010)}}
\end{array}
$ & $1\ 10^{-7}$ \\ \hline
\end{tabular}
\\ \hline\hline
\end{tabular}
\]

\begin{quotation}
Table 5. {\small Direct measurement of }$\gamma ${\small : Some space
experiments dedicated to }$\gamma ${\small .}
\end{quotation}

\end{landscape}

\end{document}